\documentclass[usenatbib,a4paper]{mn2e}

\usepackage{epsfig,times,amsmath,supertabular,colortbl,amsfonts,amssymb,multirow,array,fleqn,longtable}

\title[Twist and Turn]{Twist and Turn: Weak Lensing Image Distortions to Second Order}

\author[D. J. Bacon \& B. M. Sch\"afer]{David J. Bacon$^{1}$\thanks{E-mail: david.bacon@port.ac.uk}
\& Bj\"orn Malte Sch\"afer$^{2}$\\ 
$^1$ Institute of Cosmology and Gravitation, University of Portsmouth,
Mercantile House, Hampshire Terrace, Portsmouth, PO1 2EG, United Kingdom \\
$^2$ Institut d'Astrophysique Spatiale, Universit{\'e} de Paris XI, B{\^a}timent 120-121, Centre Universitaire d'Orsay, 91400 Orsay CEDEX, France}

\begin{document}
\maketitle

\newcommand{\expect}[1]{\left\langle #1 \right\rangle} 

\newcommand{\fflex}{\mbox{$\mathcal{F}$}}
\newcommand{\fflexsmall}{\mbox{\tiny $\mathcal{F}$}}
\newcommand{\gflex}{\mbox{$\mathcal{G}$}}
\newcommand{\gflexsmall}{\mbox{\tiny $\mathcal{G}$}}
\newcommand{\fflext}{\mbox{$\mathcal{F_T}$}}
\newcommand{\gflext}{\mbox{$\mathcal{G_T}$}}
\newcommand{\stild}{\tilde{\gamma}}
\newcommand{\ftild}{\tilde{\mbox{$\mathcal{F}$}}}
\newcommand{\gtild}{\tilde{\mbox{$\mathcal{G}$}}}
\newcommand{\mR}{\mbox{$\mathcal{R}$}}
\newcommand{\trans}{\mbox{$\mathcal{T}$}}
\newcommand{\tri}{\mathcal{T}}
\newcommand{\dif}{\mbox{$\mathrm{d}$}}
\newcommand{\x}{\mbox{\boldmath$x$}}
\newcommand{\y}{\mbox{\boldmath$y$}}
\newcommand{\thetab}{\mbox{\boldmath$\theta$}}
\newcommand{\Delthetab}{\mbox{\boldmath$\Delta\theta$}}
\newcommand{\alphab}{\mbox{\boldmath$\alpha$}}
\newcommand{\betab}{\mbox{\boldmath$\beta$}}
\newcommand{\xib}{\mbox{\boldmath$\xi$}}
\newcommand{\nablab}{\mbox{\boldmath$\nabla$}}
\newcommand{\me}{\mbox{$\mathrm{e}$}}
\newcommand{\mi}{\mbox{$\mathrm{i}$}}
\newcommand{\mes}{\mbox{\scriptsize$\mathrm{e}$}}
\newcommand{\mos}{\mbox{\scriptsize$\mathrm{o}$}}
\newcommand{\mess}{\mbox{\tiny$\mathrm{e}$}}
\newcommand{\moss}{\mbox{\tiny$\mathrm{o}$}}
\newcommand{\msep}{\mbox{\scriptsize$\mathrm{sep}$}}
\newcommand{\mis}{\mbox{\scriptsize$\mathrm{i}$}}
\newcommand{\ml}{\mbox{\scriptsize$\mathrm{l}$}}
\newcommand{\mls}{\mbox{\scriptsize$\mathrm{ls}$}}
\newcommand{\ms}{\mbox{\scriptsize$\mathrm{s}$}}
\newcommand{\mE}{\mbox{\scriptsize$\mathrm{E}$}}
\newcommand{\mB}{\mbox{\scriptsize$\mathrm{B}$}}
\newcommand{\mc}{\mbox{\scriptsize$\mathrm{c}$}}
\newcommand{\mH}{\mbox{$\mathrm{H}$}}
\newcommand{\mkms}{\mbox{$\mathrm{kms}$}}
\newcommand{\mkg}{\mbox{$\mathrm{kg}$}}
\newcommand{\mMpc}{\mbox{$\mathrm{Mpc}$}}
\newcommand{\mcov}{\mbox{$\mathrm{cov}$}}
\newcommand{\mN}{\mbox{\scriptsize$\mathrm{N}$}}
\newcommand{\mint}{\mbox{\scriptsize$\mathrm{int}$}}
\newcommand{\mcrit}{\mbox{\scriptsize$\mathrm{crit}$}}
\newcommand{\mtot}{\mbox{\scriptsize$\mathrm{tot}$}}
\newcommand{\mrot}{\mbox{\scriptsize$\mathrm{rot}$}}
\newcommand{\mprop}{\mbox{\scriptsize$\mathrm{prop}$}}
\newcommand{\mcom}{\mbox{\scriptsize$\mathrm{com}$}}
\newcommand{\mang}{\mbox{\scriptsize$\mathrm{ang}$}}
\newcommand{\mlum}{\mbox{\scriptsize$\mathrm{lum}$}}
\newcommand{\mvir}{\mbox{\scriptsize$\mathrm{vir}$}}
\newcommand{\mstar}{\mbox{\scriptsize$\mathrm{star}$}}
\newcommand{\mobs}{\mbox{\scriptsize$\mathrm{obs}$}}
\newcommand{\mMC}{\mbox{\scriptsize$\mathrm{MC}$}}
\newcommand{\mMCs}{\mbox{\tiny$\mathrm{MC}$}}
\newcommand{\atanh}{\mbox{$\mathrm{tanh}$}}
\newcommand{\nn}{\nonumber \\}
\newcommand{\sextractor}{\textsc{SExtractor}}
\newcommand{\mmin}{\mbox{\scriptsize$\mathrm{min}$}}
\newcommand{\mmax}{\mbox{\scriptsize$\mathrm{max}$}}
\newcommand{\munweighted}{\mbox{\scriptsize$\mathrm{unweighted}$}}
\newcommand{\mGaussian}{\mbox{\scriptsize$\mathrm{Gauss}$}}
\newcommand{\mdiag}{\mbox{\scriptsize$\mathrm{diag}$}}
\newcommand{\mmag}{\mbox{\scriptsize$\mathrm{unweighted}$}}
\newcommand{\mav}{\mbox{\scriptsize$\mathrm{av}$}}
\newcommand{\mshot}{\mbox{\scriptsize$\mathrm{shot}$}}
\newcommand{\mlin}{\mbox{\scriptsize$\mathrm{lin}$}}
\newcommand{\mback}{\mbox{\scriptsize$\mathrm{back}$}}
\newcommand{\mcoll}{\mbox{\scriptsize$\mathrm{coll}$}}
\newcommand{\mf}{\mbox{\scriptsize$\mathrm{f}$}}
\newcommand{\mhalos}{\mbox{\scriptsize$\mathrm{halos}$}}
\newcommand{\mm}{\mbox{\scriptsize$\mathrm{m}$}}
\newcommand{\lcdm}{$\Lambda$CDM}
\newcommand{\ard}{\hat{a}^{\dagger}_r}
\newcommand{\ald}{\hat{a}^{\dagger}_l}
\newcommand{\ards}{\hat{a}^{\dagger 2}_r}
\newcommand{\alds}{\hat{a}^{\dagger 2}_l}
\newcommand{\ar}{\hat{a}_r}
\newcommand{\al}{\hat{a}_l}
\newcommand{\ars}{\hat{a}^2_r}
\newcommand{\als}{\hat{a}^2_l}

\newcommand{\nat}{Nat}
\newcommand{\mnras}{MNRAS}
\newcommand{\apj}{ApJ}
\newcommand{\apjl}{ApJL}
\newcommand{\apjs}{ApJS}
\newcommand{\physrep}{Phys.~Rep.}
\newcommand{\aap}{A\&A}
\newcommand{\aaps}{A\&AS}
\newcommand{\aj}{AJ}
\newcommand{\prd}{Phys.~Rev.~D.}

\begin{abstract}
We account for all the image distortions relevant to weak gravitational lensing to second order. Besides the familiar shear, convergence, rotation and flexions, we find a new image distortion with two distinct descriptions, the twist and the turn. Like rotation, this distortion is not activated gravitationally to first order, but will be activated by systematic effects. We examine the rotational properties of twist and turn, and their effect on images in real and shapelet space. We construct estimators for the new distortion, taking into account the centroid shift which it generates. We then use these estimators to make first constraints on twist using the STAGES HST survey; we find that the mean twist estimator is consistent with zero. We measure correlation functions for our twist estimator on the survey, again finding no evidence of systematic effects.
\end{abstract}


\section{Introduction}

Weak gravitational flexion is a relatively new addition to the panoply of gravitational lensing effects, but has considerable potential for measuring substructure in the density distribution of matter in the Universe \citep[see e.g.][]{2002ApJ...564...65G, 2005ApJ...619..741G, 2005NewAR..49...83I, 2006MNRAS.365..414B, 2008ApJ...680....1O, 2008A&A...485..363S}. 

Flexion is proportional to third angular derivatives of the projected gravitational potential along the line of sight. As such, it is at the next order of differentiation compared to shear and convergence, which are the more studied weak lensing measures \citep[see][ for an extensive review]{2001PhR...340..291B}. Since, as we shall see, there are two independent combinations of third derivatives, there are two different flexion effects: the 1-flexion, which is a vector distortion leading to objects being skewed; and the 3-flexion, which is a spin three distortion changing circular objects into trefoils.

Up until now, these have been the only known image distortions at this order. However, in this paper we will show that there is a further neglected image distortion at the flexion level, with two alternative descriptions which we call twist and turn for reasons which will become obvious. This distortion is not activated by gravity under the most straightforward approximations; but it will be activated by systematic effects. The latter are of great concern to weak lensing, so finding a further signature of systematics is potentially very valuable to upcoming lensing surveys.

In this paper we show how twist or turn arises, and account for why it has not been noticed before. We show how it affects images in real and shapelet space, and give details of how it can be measured with fairly straightforward estimators. We then measure twist for the first time using the Space Telescope A901/902 Galaxy Evolution Survey \citep[STAGES,][]{2007AAS...21113220G}, a large mosaic observed with the Hubble Space Telescope (HST). We show that the twist is consistent with zero for STAGES on all scales, both in terms of its mean values and its correlation functions, incrementally adding confidence in the management of systematics for this survey.

The paper is organised as follows. In Section~\ref{sect_lensing_first_order}, we recount the theory of image distortions in weak lensing at the more studied first order. We note that there is already a non-gravitational mode at this order; image rotation. We write the distortions in terms of Pauli matrices, which will give us the necessary clues for how to treat higher order distortions later. 

In Section~\ref{sect_lensing_second_order}, we extend the account to second order. We find that there are combinations of Pauli matrices orthogonal to those describing the conventional flexion degrees of freedom; these orthogonal combinations give twist and turn distortions. We are therefore able to write down for the first time the complete weak image distortion to second order, and show how twist and turn are related to one another observationally.

Section~\ref{sect_twist_turn} describes the behaviour of twist/turn. The rotational properties of the distortion is worked out, and we find that is is a vector quantity. We show the impact of twist and turn on simple images; we find that they do not affect the shape of circularly symmetric images, but only images with non-zero ellipticity. We show explicitly the nature of twist and turn in shapelet space, proving that they have no impact on circularly symmetric sources, and derive how they move power between shapelet coefficients. 

In Section~\ref{sect_estimators} we go about finding practical estimators for measuring twist and turn. We derive simple estimators in shapelet space. Noting that like flexion, twist and turn affect the centroids of objects, we correct the estimators by constructing slightly more complicated expressions which take this shift into account. However, we will show that our estimators are not perfect; they would respond to flexion if it is present, and should therefore be treated as estimators of any second-order systematics, or of (real or systematic) twist on scales where flexion is negligible.

In Section~\ref{sect_measurement} we use these estimators to constrain twist observationally for the first time, using the STAGES HST survey. We find that our twist estimator has a larger variance than flexion estimators, and that its mean value is consistent with zero in the STAGES data. We measure correlation functions for twist estimators, again finding that they are consistent with zero systematic in STAGES. We summarise our results and conclude in Section~\ref{sect_summary}.

\section{Image Distortions to First Order}\label{sect_lensing_first_order}

We begin by discussing image distortions in weak lensing to first order \citep[for more details, see][]{2001PhR...340..291B}. We can describe the effect of lensing as a mapping between the surface brightness $f_S$ of a galaxy at a position $(\beta_1,\beta_2)$ in the source plane, and the surface brightness $f_I$ at a position $(\theta_1,\theta_2)$ in the image plane:
\begin{equation}
f_I(\theta_i)=f_S(\beta_i)=f_S(A_{ij} \theta_j)
\end{equation}
where we have set the origin of $\theta_i$ and $\beta_i$ to the centre of light in the respective planes. $A$ is the Jacobian matrix which maps image positions to source positions,
\begin{equation}
A_{ij}=\frac{\partial \beta_i}{\partial \theta_j}.
\end{equation}
For lensing with a single lens plane, and assuming the Born approximation, this is given by
\begin{equation}
A_{ij}=\delta_{ij}-\partial_i\partial_j \psi
\end{equation}
where $\psi$ is the lensing potential, i.e. the gravitational potential suitably projected into 2D. We can therefore write $A$ as
\begin{equation}
A=\left(\begin{array}{cc}1-\kappa&0\\0&1-\kappa\end{array}\right)+\left(\begin{array}{cc}-\gamma_1&-\gamma_2\\-\gamma_2&\gamma_1\end{array}\right)
\end{equation}
with the convergence $\kappa$ given by
\begin{equation}
\kappa=\frac{1}{2}(\partial_1^2+\partial_2^2)\psi
\end{equation}
and the shear $\gamma_i$ given by
\begin{equation}
\gamma_1=\frac{1}{2}(\partial_1^2-\partial_2^2)\psi, \hspace{1cm} \gamma_2=\partial_1\partial_2\psi.
\end{equation}
There is an alternative notation that is useful to us, introduced by \citet{2006MNRAS.365..414B}. We define the complex derivative $\partial\equiv\partial_1+i\partial_2$; in cylindrical coordinates this is given by
\begin{equation}
\partial={\rm e}^{i\phi}\left(\frac{\partial}{\partial \theta}+\frac{i}{\theta}\frac{\partial}{\partial\phi}\right)
\label{eq:spin}
\end{equation}
with radial coordinate $\theta$ and azimuthal coordinate $\phi$. We also define $\gamma\equiv\gamma_1+i\gamma_2$, and then 
\begin{equation}
\kappa=\frac{1}{2}\partial\partial^*\psi, \hspace{1cm} \gamma=\frac{1}{2}\partial\partial\psi.
\label{eq:kappagamma}
\end{equation}
Besides simplifying notation, this format elucidates the spins of the quantities; when $\partial$ is applied, the e$^{i\phi}$ term in equation (\ref{eq:spin}) raises the spin by one. Similarly,
the application of $\partial^*$ lowers the spin by one. So since $\psi$ is a scalar, so is $\kappa$, while $\gamma$ is spin 2.

However, our study of $A$ is not complete. We have specified three quantities in $A$, i.e. $\kappa, \gamma_1$, and $\gamma_2$. But $A$ is a four element object, so there is a further degree of freedom which we have missed. We quickly realise that this is a {\bf rotation} $\rho$, i.e.
\begin{equation}
A=\left(\begin{array}{cc}1-\kappa&0\\0&1-\kappa\end{array}\right)+\left(\begin{array}{cc}-\gamma_1&-\gamma_2\\-\gamma_2&\gamma_1\end{array}\right)+\left(\begin{array}{cc}0&\rho\\-\rho&0\end{array}\right)
\end{equation}
for small rotation angles $\rho$. Whereas $\kappa$ and $\gamma$ can be written as second derivatives of the lensing potential, this is not possible for $\rho$. It is not activated by gravity in our approximation (due to the interchangability of the second derivatives of the gravitational potential, $\partial_i\partial_j\psi = \partial_j\partial_i\psi$), but may be present in a real lensing survey as a systematic \citep[see e.g. the rotation caused by the telescope constrained by][]{2000MNRAS.318..625B}. This rotation has been described previously; see e.g. \citet{2003PhRvD..68h3002H}.

We will find it convenient to write $A$ as a sum of Pauli matrices, as these provide an orthogonal basis for studying further degrees of freedom at the next order of weak lensing approximation. The Pauli matrices are given by \citep{2005mmp..book.....A}
\begin{eqnarray}
\nonumber I=\left(\begin{array}{cc}1&0\\0&1\end{array}\right)&\hspace{.5cm}&\sigma_1=\left(\begin{array}{cc}0&1\\1&0\end{array}\right)\\
\sigma_2=\left(\begin{array}{cc}0&-i\\i&0\end{array}\right)&\hspace{.5cm}&\sigma_3=\left(\begin{array}{cc}1&0\\0&-1\end{array}\right)
\end{eqnarray}
so we can write $A$ as
\begin{equation}
A=(1-\kappa)I-\gamma_1\sigma_3-\gamma_2\sigma_1+\rho i\sigma_2.
\label{eq:pauli}
\end{equation}
We will need one further concept: in a weak lensing context, it is usual to assume that the shear and convergence are small and constant across an object. We can then write the surface brightness mapping as
\begin{equation}
f_I(\theta_i)=f_S(\delta_{ij}\theta_j+(A_{ij}-\delta_{ij})\theta_j)\simeq f_S(\theta_i)+(A_{ij}-\delta_{ij})\theta_j\partial_i f_S(\theta_i)
\end{equation}
We will now modify this to show how flexion and the new distortions enter.

\section{Image Distortions to Second Order}\label{sect_lensing_second_order}

The further step taken by flexion studies is to note that in reality, shear will vary across an object. If we keep $A$ as a constant across the object, we need a further term in a Taylor expansion in the surface brightness map, as given by \citet{2005ApJ...619..741G},
\begin{equation}
f_I(\theta_i)=f_S\left(A_{ij}\theta_j + \frac{1}{2}D_{ijk}\theta_j\theta_k\right).
\label{eq:f2nd}
\end{equation}
This introduces the $D$ tensor; if we suppose that its components are purely due to a variation of $A$ across the image, we can write $D_{ijk}=\partial_k A_{ij}$. Then by differentiating equation (\ref{eq:pauli}) we find
\begin{eqnarray}
D_{ij1}=-\partial_1\kappa I-\partial_1\gamma_1\sigma_3-\partial_1\gamma_2\sigma_1+\partial_1\rho i\sigma_2,\nonumber\\
D_{ij2}=-\partial_2\kappa I-\partial_2\gamma_1\sigma_3-\partial_2\gamma_2\sigma_1+\partial_2\rho i\sigma_2.
\end{eqnarray}
We can rewrite much of this in terms of flexion. We define the 1-flexion as $F\equiv F_1+iF_2$, and the 3-flexion $G\equiv G_1+iG_2$, where
\begin{equation}
F=\frac{1}{2}\partial\partial\partial^*\psi, \hspace{1cm} G=\frac{1}{2}\partial\partial\partial\psi.
\end{equation}
$F$ is manifestly spin 1 and $G$ is spin 3. Comparing with equation (\ref{eq:kappagamma}) and disentangling the individual components we find
\begin{eqnarray}
F_1&=&\partial_1\kappa=\partial_1\gamma_1+\partial_2\gamma_2,\nonumber\\
F_2&=&\partial_2\kappa=\partial_1\gamma_2-\partial_2\gamma_1,\nonumber\\
G_1&=&\partial_1\gamma_1-\partial_2\gamma_2,\nonumber\\
G_2&=&\partial_1\gamma_2+\partial_2\gamma_1.
\end{eqnarray}
Reorganising in terms of derivatives of shear, we can write
\begin{eqnarray}
\partial_1\gamma_1&=&\frac{1}{2}(F_1+G_1),\nonumber\\
\partial_2\gamma_2&=&\frac{1}{2}(F_1-G_1),\nonumber\\
\partial_1\gamma_2&=&\frac{1}{2}(F_2+G_2),\nonumber\\
\partial_2\gamma_1&=&\frac{1}{2}(-F_2+G_2).
\end{eqnarray}
Hence we can write the $D$ tensor in terms of the Pauli matrices as
\begin{eqnarray}
D_{ij1}=-F_1 I-\frac{1}{2}(F_1+G_1)\sigma_3-\frac{1}{2}(F_2+G_2)\sigma_1+\frac{1}{2}C_1 i\sigma_2\nonumber\\
D_{ij2}=-F_2 I+\frac{1}{2}(F_2-G_2)\sigma_3-\frac{1}{2}(F_1-G_1)\sigma_1+\frac{1}{2}C_2 i\sigma_2
\end{eqnarray}
where we have defined the {\bf turn}, 
\begin{equation}
C_i=2\partial_i\rho
\end{equation}
which in the complex notation we can write as $C=2\partial\rho$, with $C\equiv C_1+i C_2$.
This is a new distortion mode, which simply describes how the amount of image rotation in the Jacobian varies across the object. Like the rotation, it is not expected to be activated by gravity at our level of approximation.

Separating into individual distortion components, we have
\begin{eqnarray}
[-2D_{ij1},-2D_{ij2}]&=&F_1 [2I+\sigma_3,\sigma_1]+F_2[\sigma_1,2I-\sigma_3]\nonumber\\
&&+G_1[\sigma_3,-\sigma_1]+G_2[\sigma_1,\sigma_3]\nonumber\\&&+C_1[-i\sigma_2,0]+C_2[0,-i\sigma_2].
\end{eqnarray}
$F$, $G$ and $C$ provide six parameters for $D$. However, $D$ has eight components, so it might initially be thought that there are eight degrees of freedom in lensing distortions at this order. What do the remaining two parameters represent? We note that whereas $C$ premultiplies $\sigma_2$, $F$ and $G$ premultiply mixtures of $I$, $\sigma_1$ and $\sigma_3$. We can therefore seek a further mixture of these latter quantities.
We can find this by writing the six known objects as 1-D lists of elements,
treating these as vectors and seeking two further vectors which are orthogonal
to these six and each other. Gaussian elimination leads to the components
\begin{eqnarray}
[-2D_{ij1},-2D_{ij2}]=...+T_1[-I+\sigma_1+\sigma_3,-I+\sigma_1-\sigma_3]\nonumber\\+T_2[-I-\sigma_1+\sigma_3,I+\sigma_1+\sigma_3].
\end{eqnarray}
where we have introduced the {\bf twist}, $T$, which might appear to be another non-gravitational distortion mode. 

However, we note from equation (\ref{eq:f2nd}) that a lensed object has surface brightness at position $\theta_i$ found using a second order term $(D_{i12}+D_{i21})\theta_1 \theta_2/2$; ie $D_{i12}$ and $D_{i21}$ do not occur independently of one another for any observational consequence. This symmetrisation means that there are six, rather than eight, observational quantities at second order. We can cause a particular distorted image by either applying a twist or a turn. We will find the relation between the two below.

\begin{figure}
\epsfig{figure=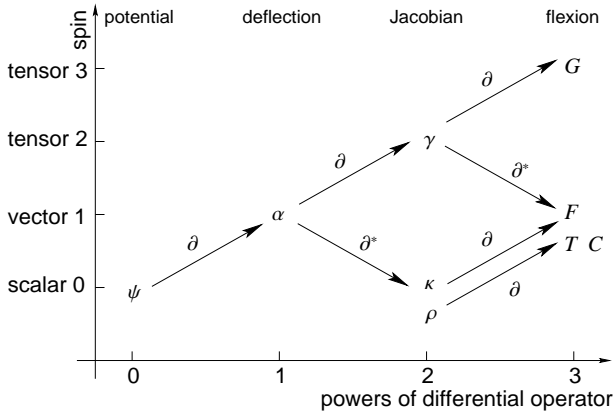,width=8cm}
\caption{Chart summarising the derivation of the lensing quantities by successive application of the differential operators, and their respective transformation properties.}
\label{figure_flexion_chart}
\end{figure}

We now have a complete list of distortions to second order. Explicitly, the full image distortion to this order is described by the $A$ matrix,
\begin{equation}
A=\left(\begin{array}{cc}1-\kappa&0\\0&1-\kappa\end{array}\right)+\left(\begin{array}{cc}-\gamma_1&-\gamma_2\\-\gamma_2&\gamma_1\end{array}\right)+\left(\begin{array}{cc}0&\rho\\-\rho&0\end{array}\right)
\end{equation}
and the $D$ tensor,
\begin{eqnarray}
-2D_{ij1}&=&\left(\begin{array}{cc}3F_1&F_2\\F_2&F_1\end{array}\right)+\left(\begin{array}{cc}G_1&G_2\\G_2&-G_1\end{array}\right)\nonumber\\&+&\left(\begin{array}{cc}0&-C_1\\C_1&0\end{array}\right)+\left(\begin{array}{cc}0&T_1-T_2\\T_1-T_2&-2T_1-2T_2\end{array}\right)\nonumber\\
-2D_{ij2}&=&\left(\begin{array}{cc}F_2&F_1\\F_1&3F_2\end{array}\right)+\left(\begin{array}{cc}G_2&-G_1\\-G_1&-G_2\end{array}\right)\nonumber\\&+&\left(\begin{array}{cc}0&-C_2\\C_2&0\end{array}\right)+\left(\begin{array}{cc}-2T_1+2T_2&T_1+T_2\\T_1+T_2&0\end{array}\right)\nonumber\\
\label{eq:d}
\end{eqnarray}
From this list of the elements of $D$ we can easily find how to convert between a twist mode and the turn which causes the same observational consequences. If we have description $a$ of a distortion with zero turn and non-zero twist $(T^a_1,T^a_2)$, this is equivalent to description $b$ with zero twist, and turn given by
\begin{equation}
\left(\begin{array}{cc}C_1^b \\ C_2^b \end{array}\right)=\left(\begin{array}{cc}1&-1\\-1&-1\end{array}\right)\left(\begin{array}{cc}T_1^a \\ T_2^a \end{array}\right).
\end{equation}
Equally if we start with description $b$, we can find description $a$ using
\begin{equation}
\left(\begin{array}{cc}T_1^a \\ T_2^a \end{array}\right)=\frac{1}{2}\left(\begin{array}{cc}1&-1\\-1&-1\end{array}\right)\left(\begin{array}{cc}C_1^b \\ C_2^b \end{array}\right).
\end{equation}
The surface brightness mapping can be approximated to second order as
\begin{eqnarray}
f_I(\theta_i)=f_S\left(\delta_{ij}\theta_j+(A_{ij}-\delta_{ij})\theta_j+\frac{1}{2}D_{ijk}\theta_j\theta_k\right)\nonumber\\
\simeq f_S(\theta_i)+(A_{ij}-\delta_{ij})\theta_j\partial_i f_S(\theta_i)+\frac{1}{2}D_{ijk}\theta_j\theta_k\partial_i f_S(\theta_i).
\label{eq:map}
\end{eqnarray}
Figure~\ref{figure_flexion_chart} gives an overview of the lensing quantities, their relationship to the gravitational potential and their transformation properties. $C$ is found by taking the derivative of $\rho$, but $\rho$ and therefore $C$ cannot be derived from the potential by taking derivatives. Thus $\rho$ and $C$ or $T$ constitute additional degrees of freedom in the lens mapping beyond gravitational effects.

\section{Behaviour of Twist and Turn}\label{sect_twist_turn}

\subsection{Rotational Properties}
We can now examine how the new constituents of $D$ transform under rotations $R$, by rotating a source-plane coordinate $\beta_i$ in the absence of $A$:
\begin{eqnarray}
\beta_i'&=&R_{il} D_{lmn} \theta_m \theta_n = R_{il}D_{lmn}R^T_{mj}R_{jp}\theta_p R^T_{nk}R_{kq}\theta_q\nonumber\\&=&R_{il}D_{lmn}R^T_{mj}R^T_{nk}\theta_j'\theta_k'
\end{eqnarray}
where primes denote rotated quantitites. But also if $D'$ is the rotated tensor then $\beta_i'=D_{ijk}' \theta_j' \theta_k' $, so
\begin{equation}
D_{ijk}'=R_{il}D_{lmn}R^T_{mj}R^T_{nk}.
\end{equation}
Since we can write the rotation by angle $\phi$ as
\begin{equation}
R=\left(\begin{array}{cc}\cos{\phi}&\sin{\phi}\\-\sin{\phi}&\cos{\phi}\end{array}\right)
\end{equation}
we can write the transformation for $D$ (and its constituent objects) as
\begin{eqnarray}
[D_{ij1}',D_{ij2}']=[R_{il}D_{lm1}R^T_{mj}\cos{\phi}+R_{il}D_{lm2}R^T_{mj}\sin{\phi},\nonumber\\-R_{il}D_{lm1}R^T_{mj}\sin{\phi}+R_{il}D_{lm2}R^T_{mj}\cos{\phi}]
\end{eqnarray}
If we set $F=G=T=C_2=0$ so that $D$ only contains non-zero turn $C_1$, we find that this $C_1$ object transforms into the equivalent $C_2$ object after a $\pi/2$ rotation, and returns to its
initial form after a $2\pi$ rotation; so $C_1$ and $C_2$ are the
components of a vector. This is to be expected, as turn is the gradient of the scalar rotation field, and so is naturally a vector.

Since a twist can be written in terms of a turn, we should expect that twist will also be a vector. Indeed, if we set $F=G=C=T_2=0$, the resulting $T_1$ object transforms into
the equivalent $T_2$ object after a $\pi/2$ rotation, and returns to its
initial form after a $2\pi$ rotation; so $T_1$ and $T_2$ are also the
components of a vector.

\subsection{Real Space Behaviour}

Now that we have established the rotational properties of the distortions, we would like to visualise what effect they have on real images. We can use the first line of the mapping equation (\ref{eq:map}) together with the $D$ tensor of equation (\ref{eq:d}) to observe the effect of the second order image distortions on a Gaussian circular or elliptical image. 

\begin{figure}
\psfig{figure=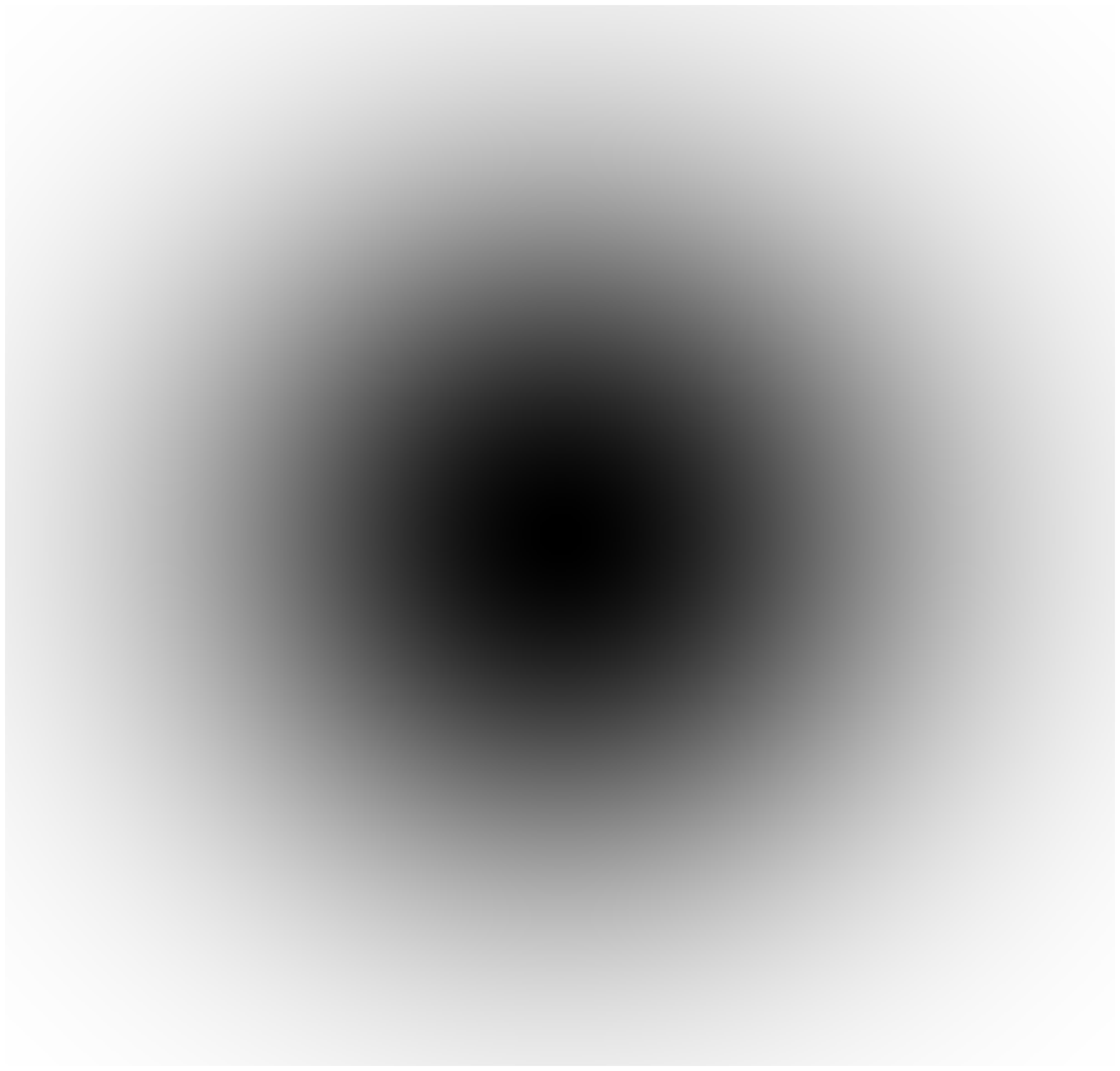,width=4cm}\psfig{figure=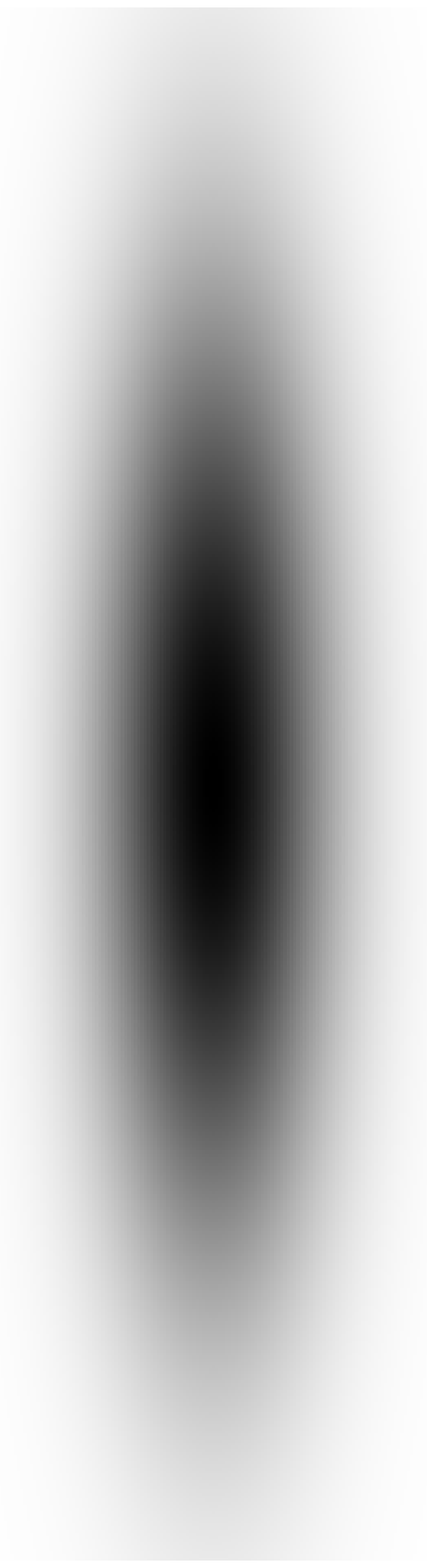,width=4cm}\vspace{-1cm}\\
\psfig{figure=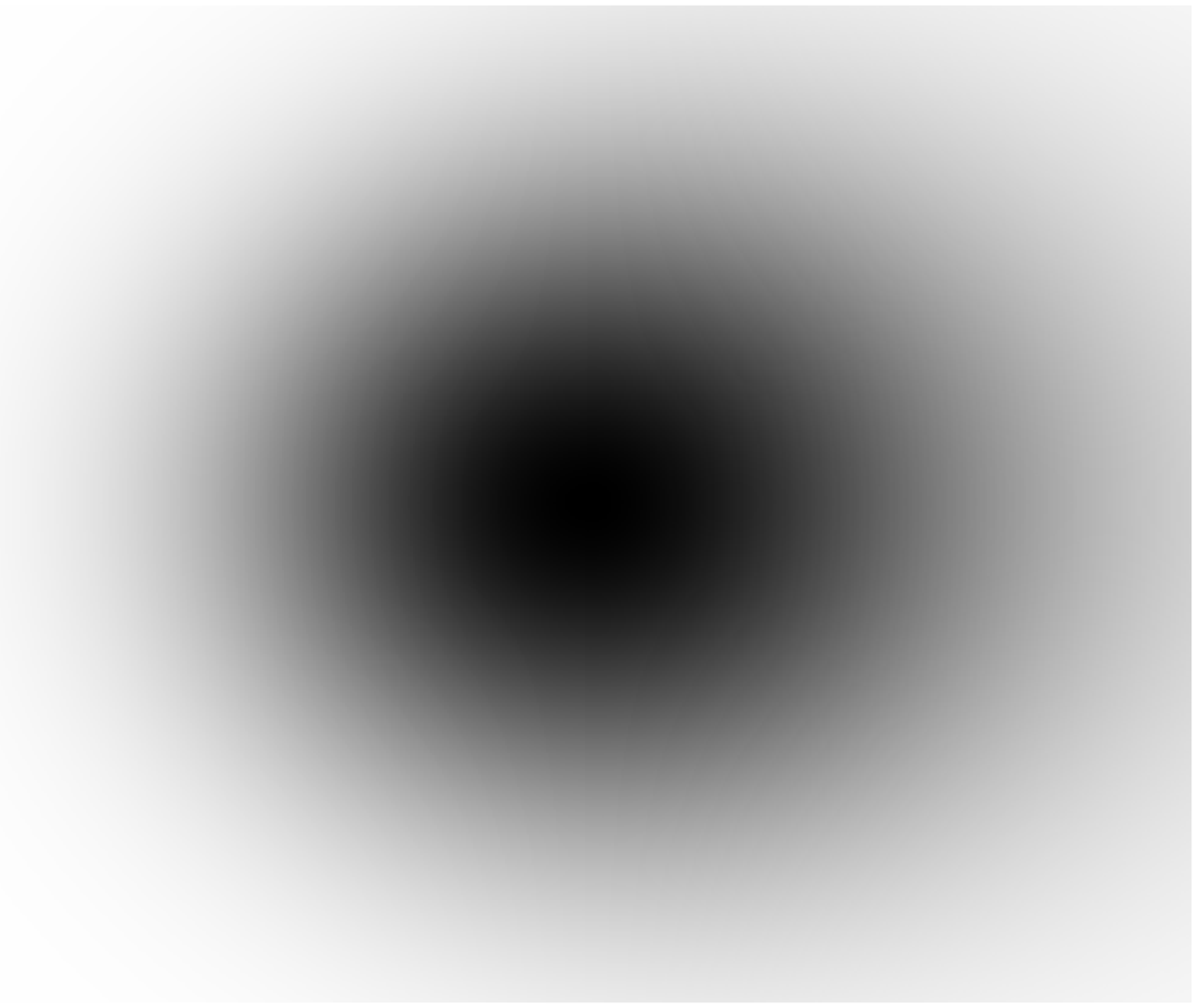,width=4cm}\psfig{figure=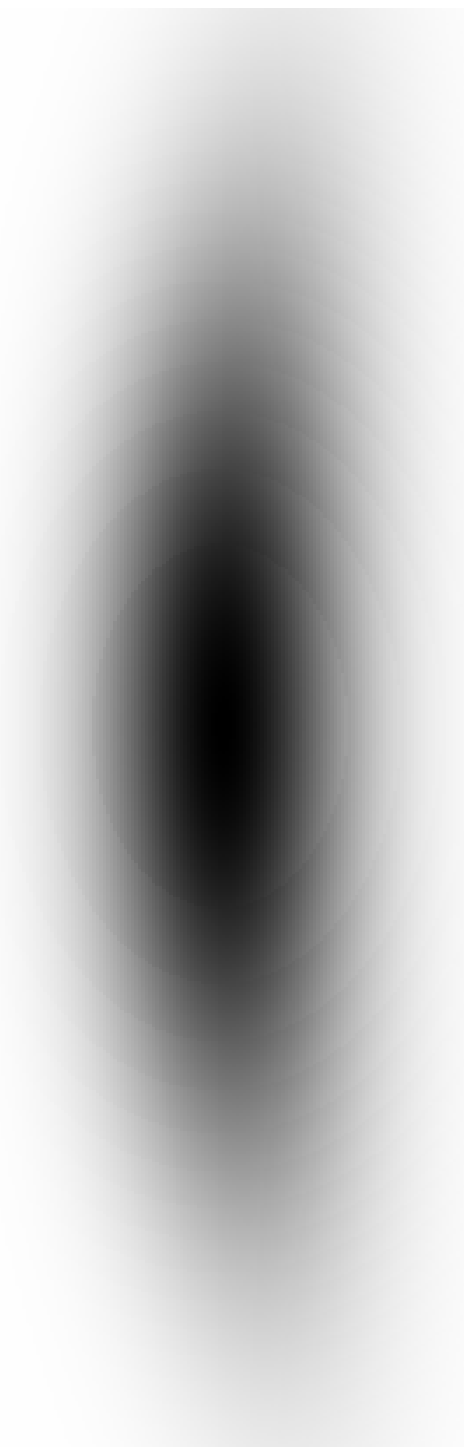,width=4cm}\vspace{-1cm}\\
\psfig{figure=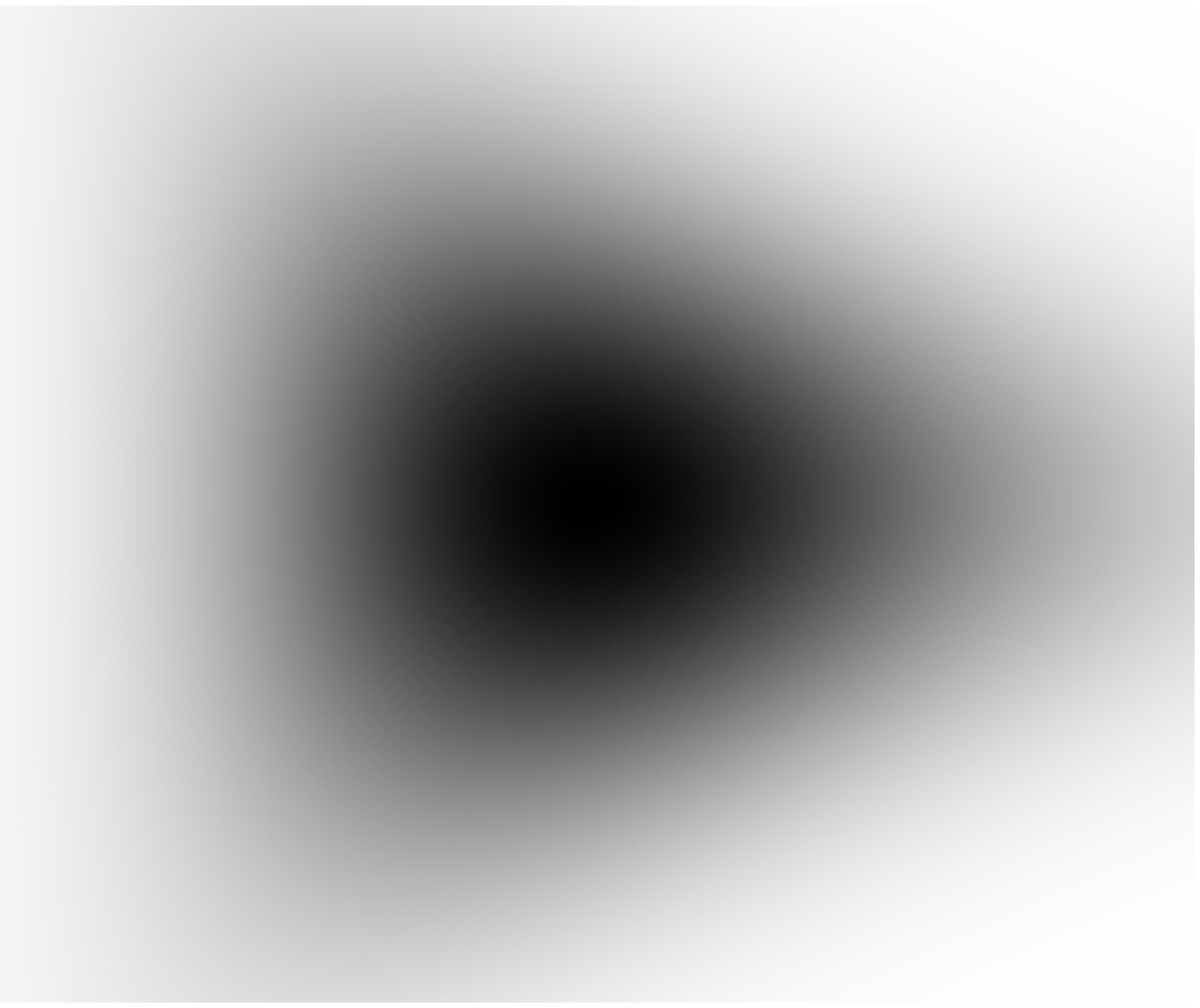,width=4cm}\psfig{figure=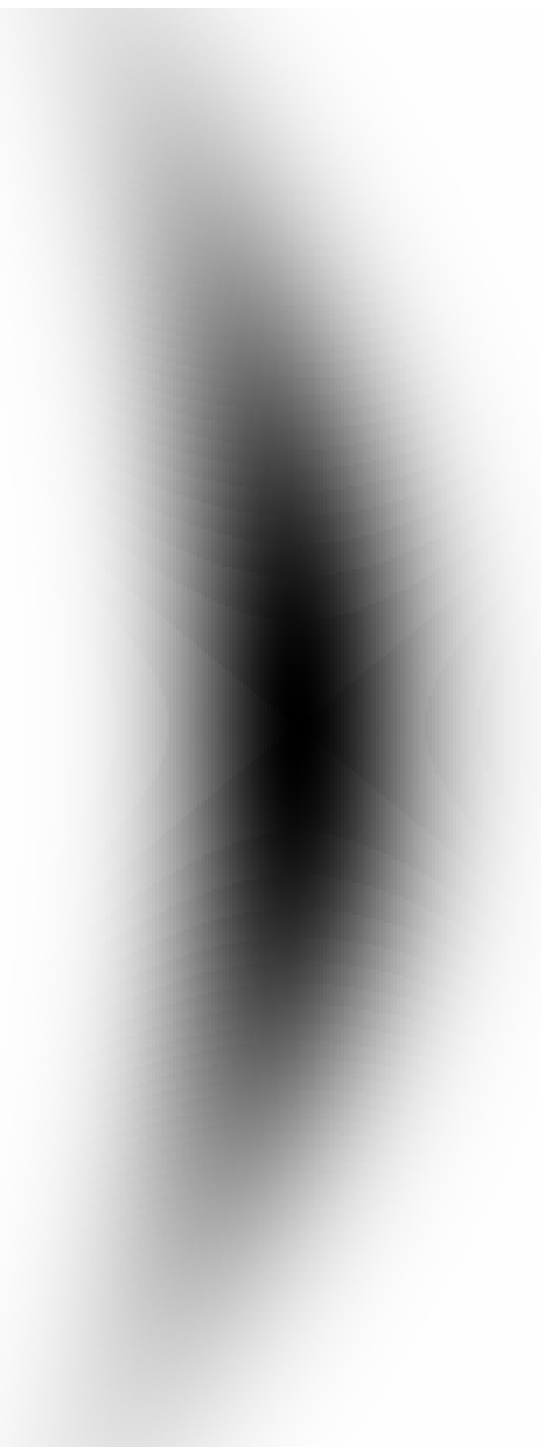,width=4cm}
\caption{Effect of 1-flexion and 3-flexion on circular and elliptical ($e=0.9$) Gaussian sources with $\sigma_{\rm major}=0.5"$. Top panel: unlensed objects; middle panel: $F_1=0.2$ arcsec$^{-1}$; bottom panel: $G_1=0.7$ arcsec$^{-1}$.}
\label{fig:flex}
\end{figure}
\begin{figure}
\center\psfig{figure=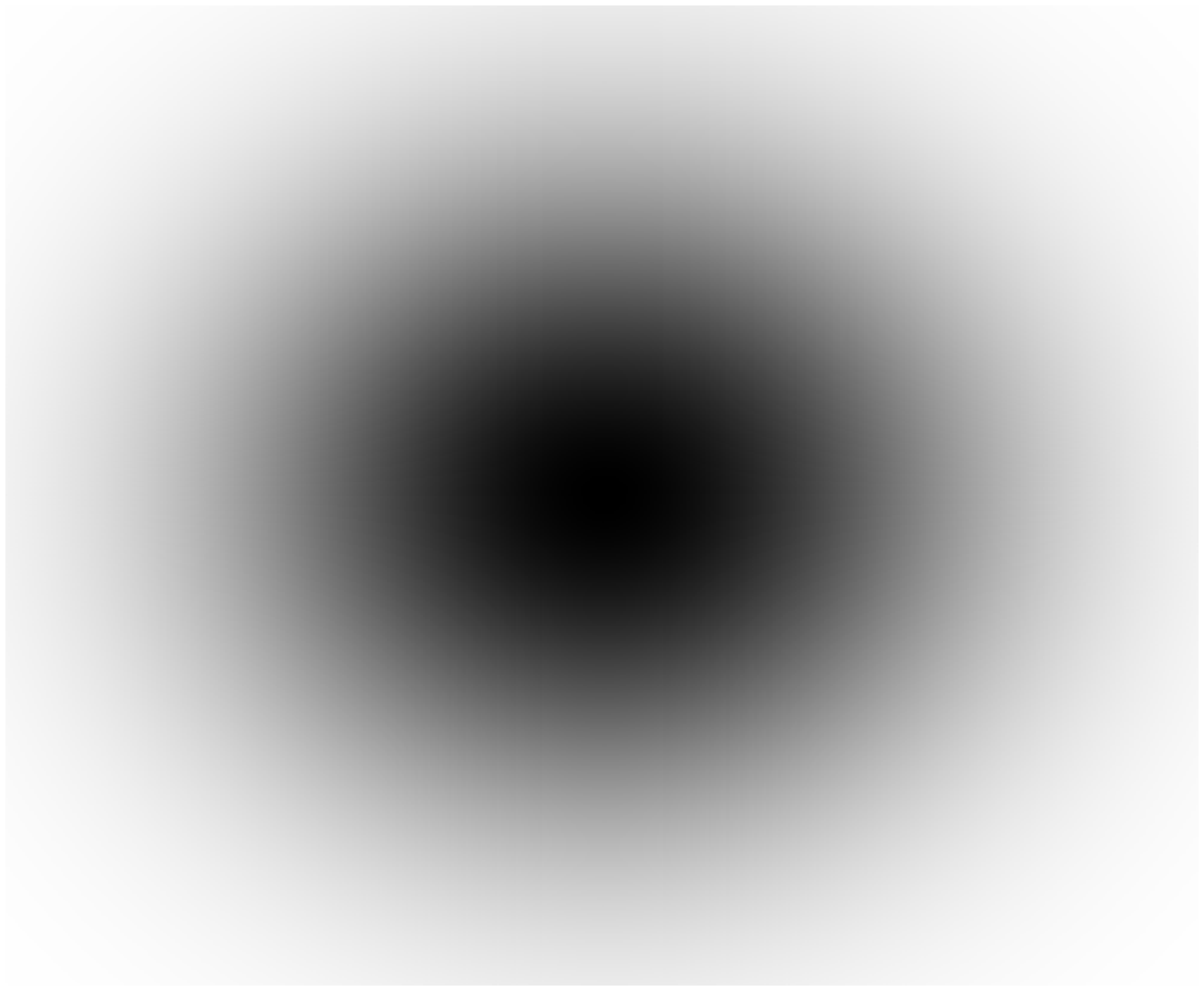,width=3.5cm}\\
\psfig{figure=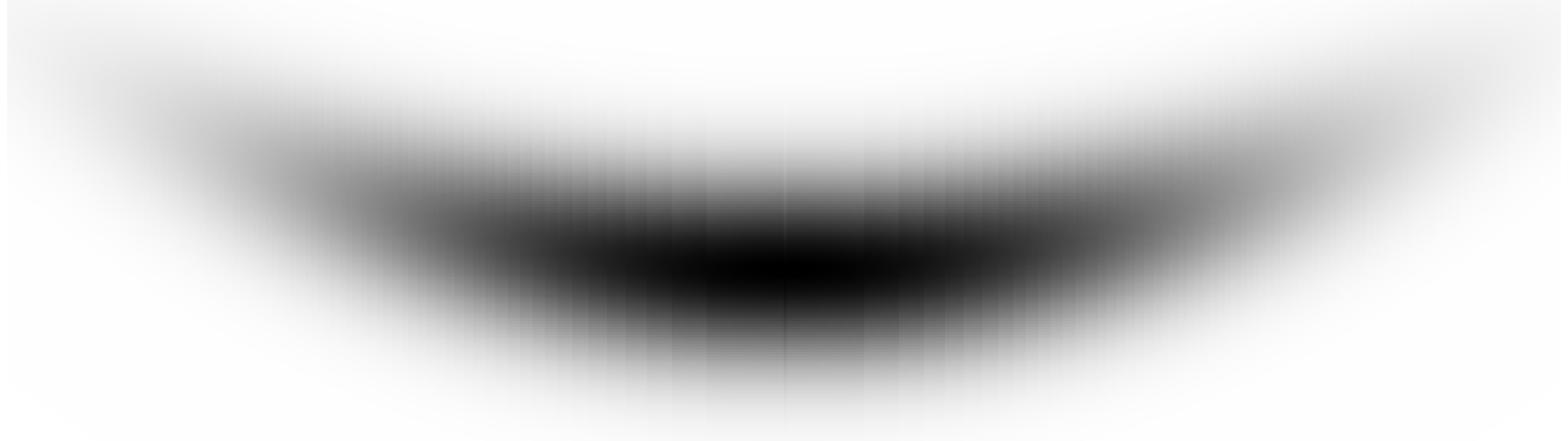,width=3.5cm}\psfig{figure=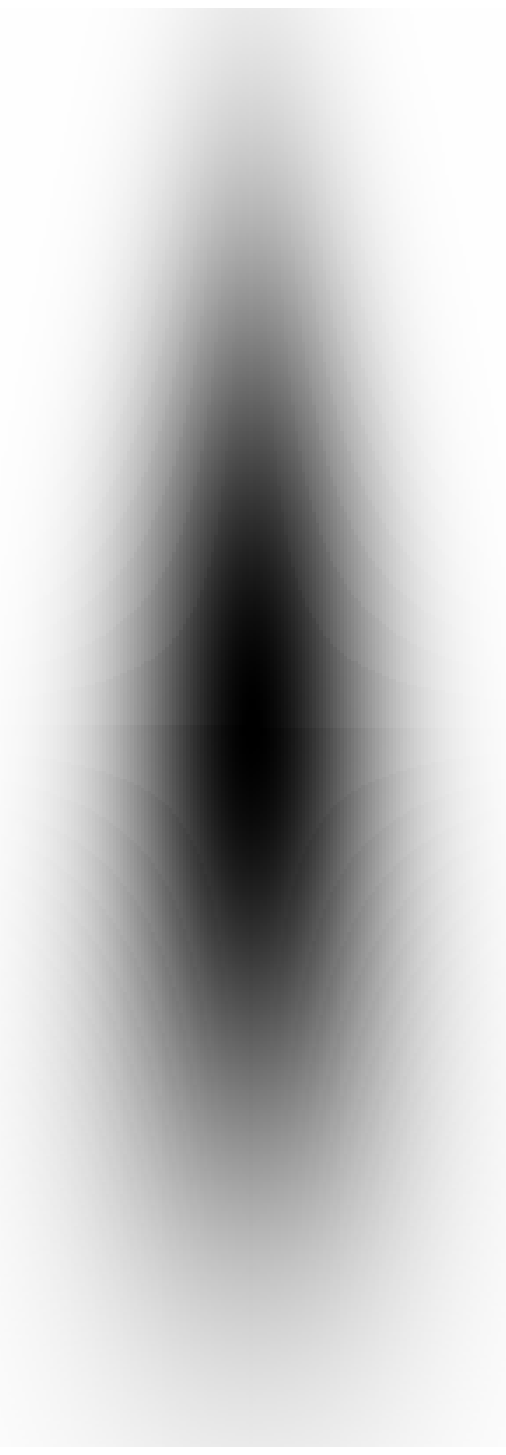,width=3.5cm}\\
\psfig{figure=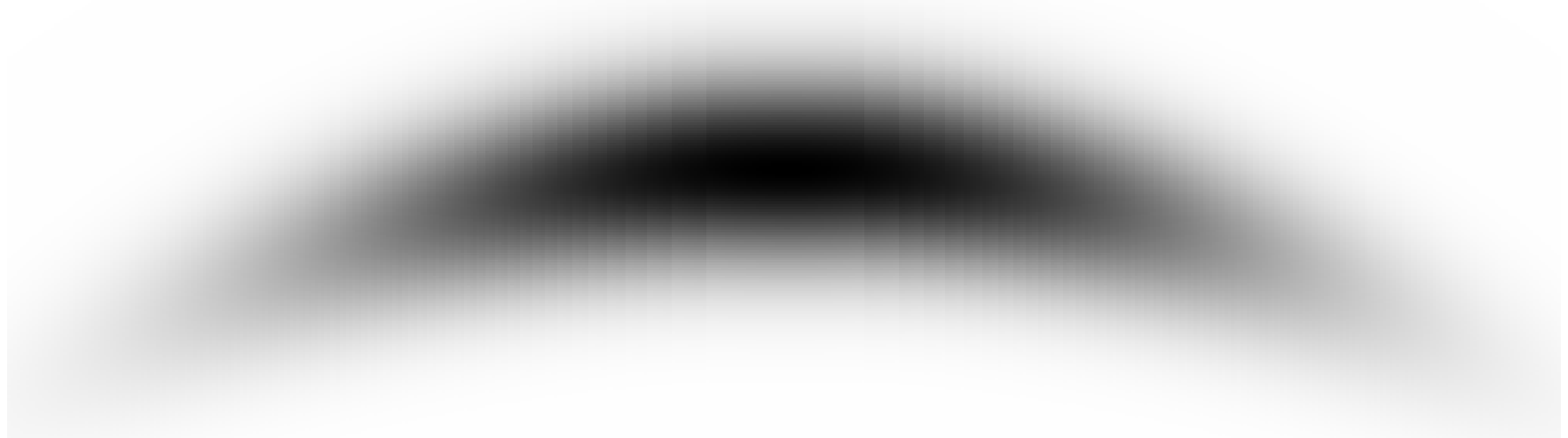,width=3.5cm}\psfig{figure=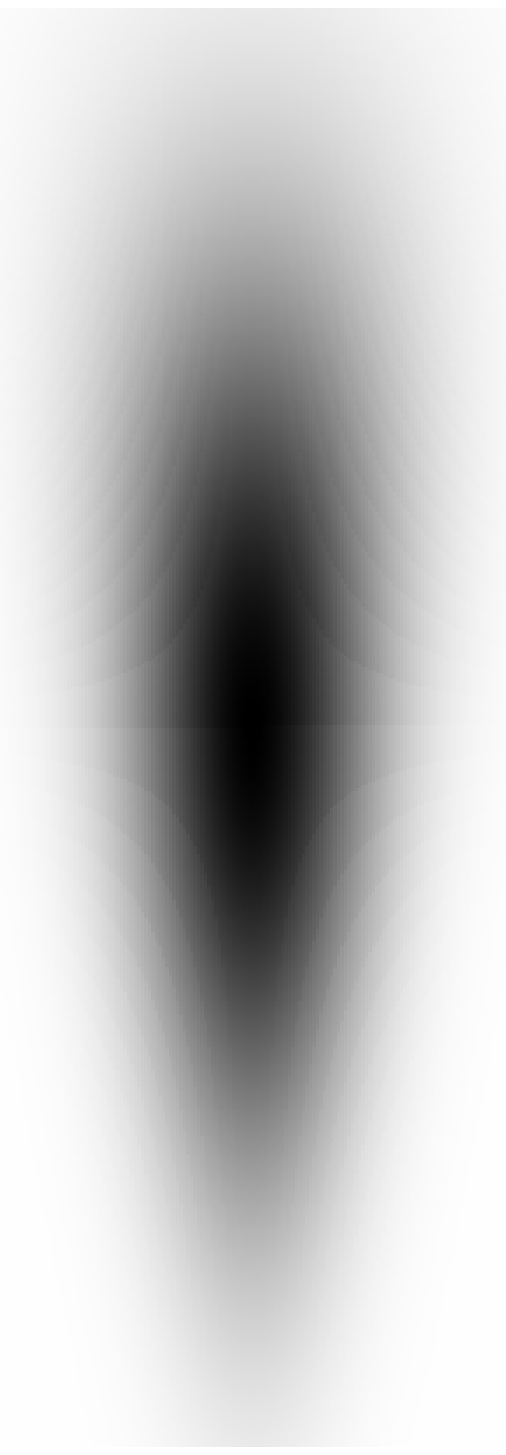,width=3.5cm}\\
\psfig{figure=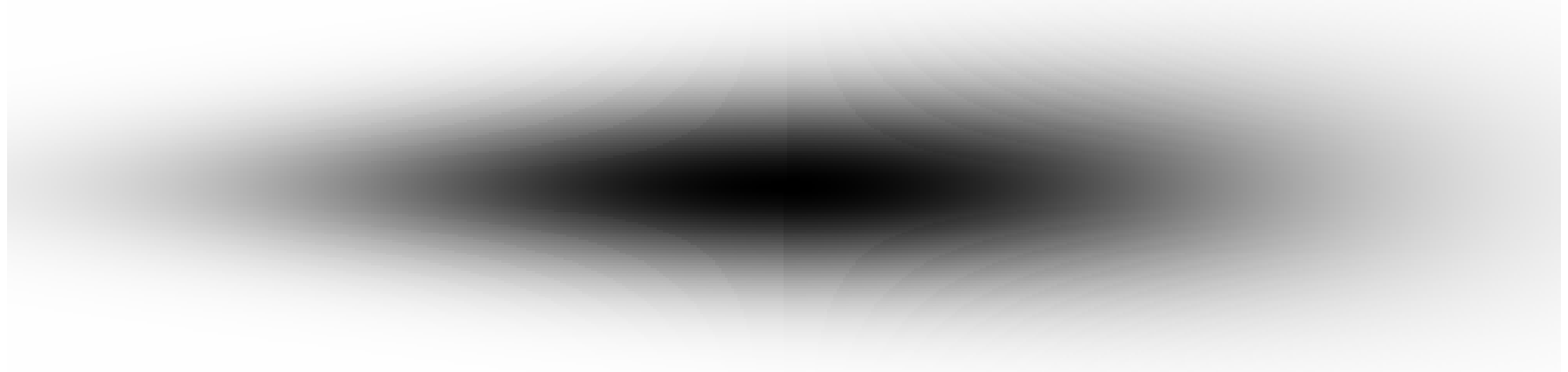,width=3.5cm}\psfig{figure=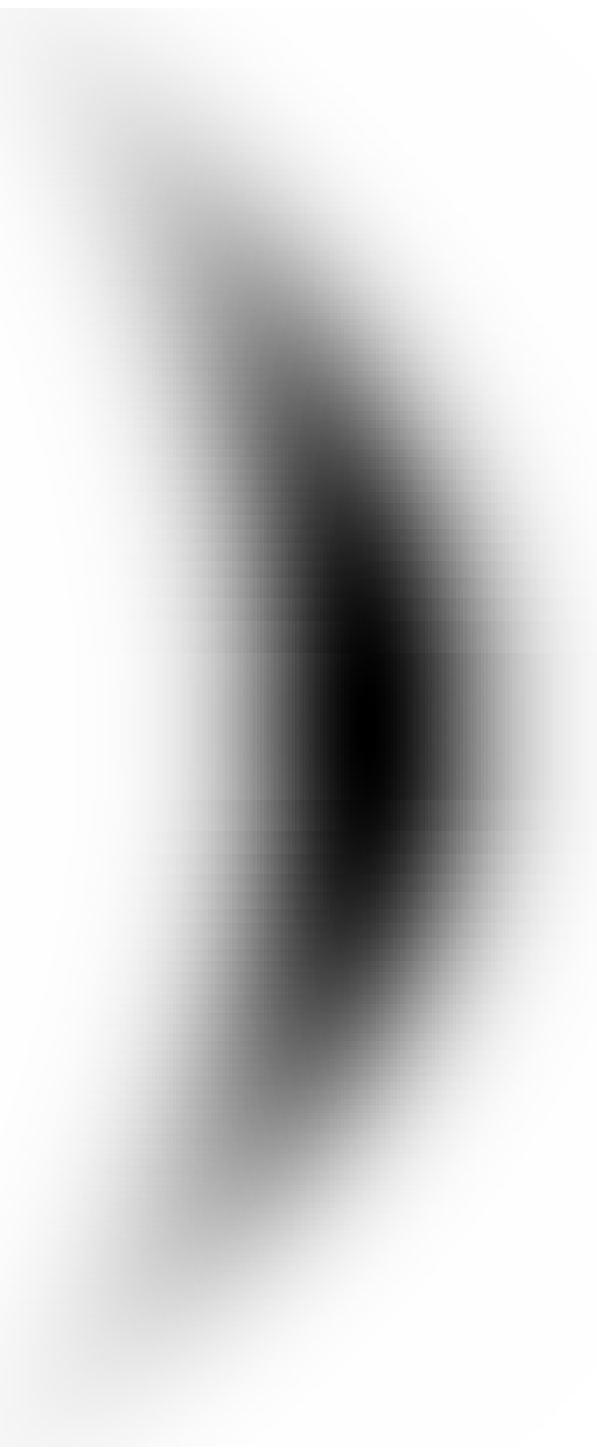,width=3.5cm}\\
\psfig{figure=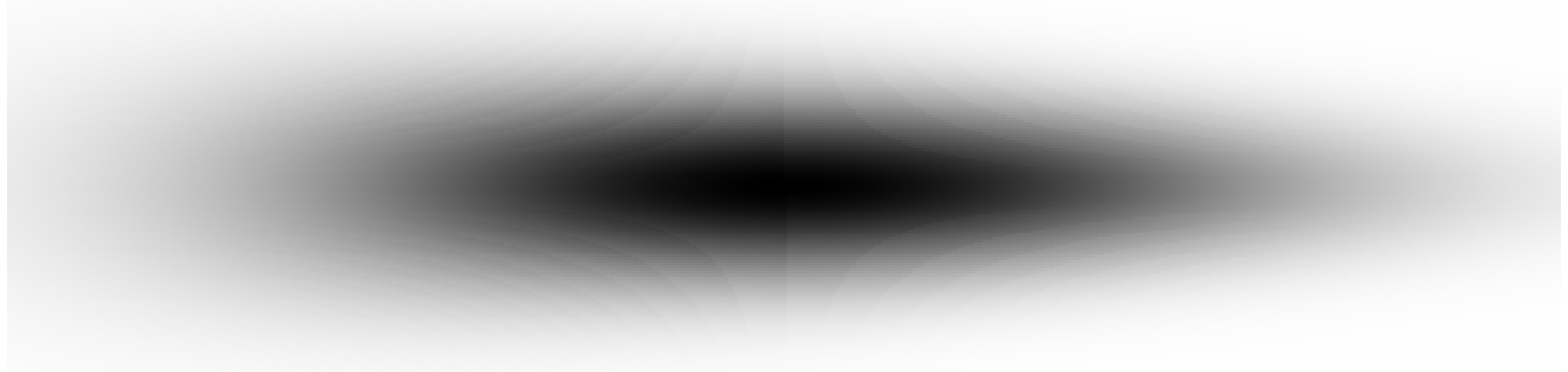,width=3.5cm}\psfig{figure=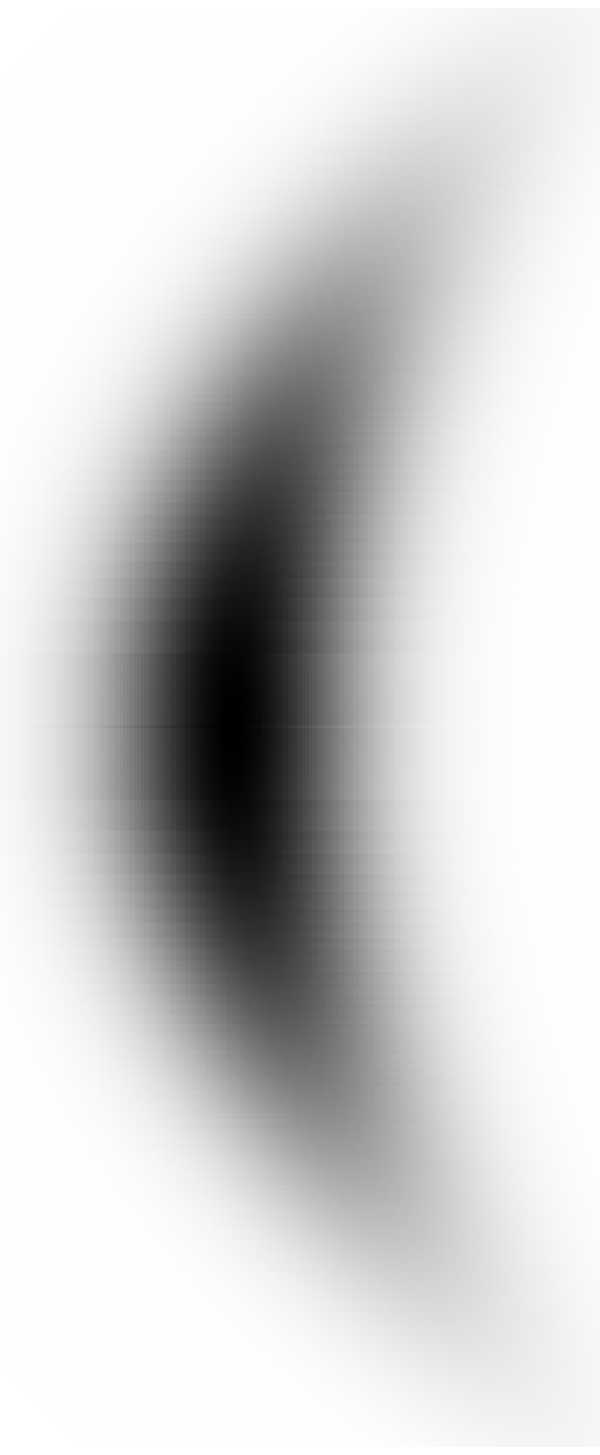,width=3.5cm}
\caption{Effect of turn on horizontally and vertically elliptical objects. Top panel: $C_1=1.3$ arcsec$^{-1}$ on a circular Gaussian with $\sigma=0.5"$. Second panel: $C_1=1.3$ arcsec$^{-1}$ on elliptical Gaussians with $\sigma_{\rm major}=0.5"$, $e_1=0.9$ and $e_1=-0.9$. Third panel: $C_1=-1.3$ arcsec$^{-1}$ on the same elliptical Gaussians. Fourth panel: $C_2=1.3$ arcsec$^{-1}$; Bottom panel: $C_2=-1.3$ arcsec$^{-1}$. }
\label{fig:turn}
\end{figure}
Figure~\ref{fig:flex} shows the effect of 1-flexion and 3-flexion. Here we have operated on objects with Gaussian surface brightness
\begin{equation}
I({\mathbf x})= A\exp{\left[-\frac{(x-x_c)^2}{2\sigma_x^2}-\frac{(y-y_c)^2}{2\sigma_y^2}\right]}
\end{equation}
where $\sigma_x=\sigma_y=0.5"$ for the circular source, and $\sigma_x=0.11", \sigma_y=0.5"$  for the elliptical source. This gives an ellipticity $e=(\sigma_y^2-\sigma_x^2)/(\sigma_y^2+\sigma_x^2)=0.9$.

Notice that 1-flexion and 3-flexion affect the shape of both circular and elliptical objects in the figure. We can compare this to figure~\ref{fig:turn}, which shows the effect of turn on circular and elliptical objects. Notice that turn has no discernible effect on the circularly symmetric source; we will show later that it indeed has zero effect on such a source. On horizontally or vertically elliptical objects, a pure $C_1$ or $C_2$ gives an arc which by eye appears similar to the impact of flexion, but it is truly a different mode of curvature with a distinct estimator which we will find below.

\begin{figure}
\center\psfig{figure=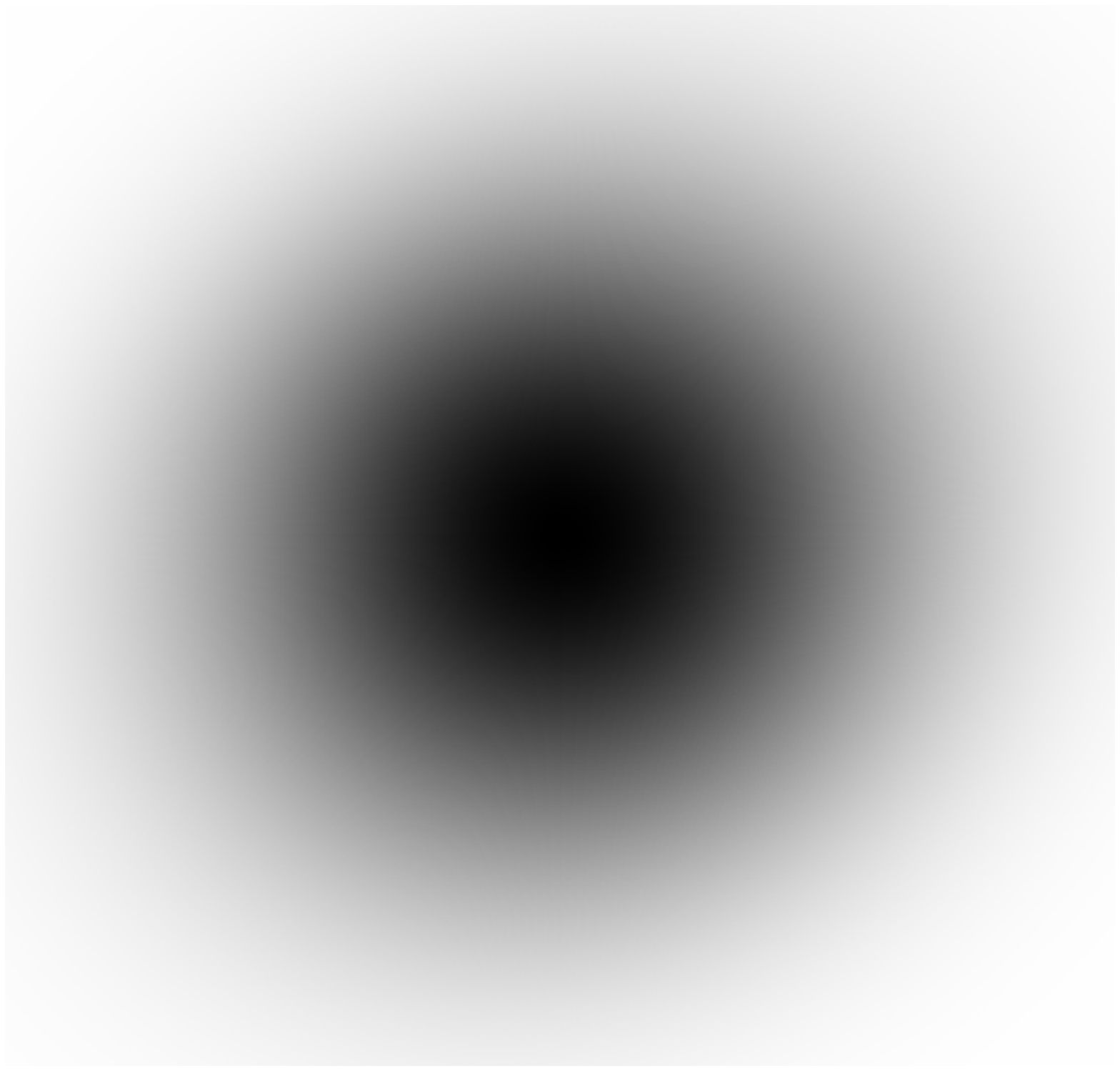,width=3.5cm}\\
\psfig{figure=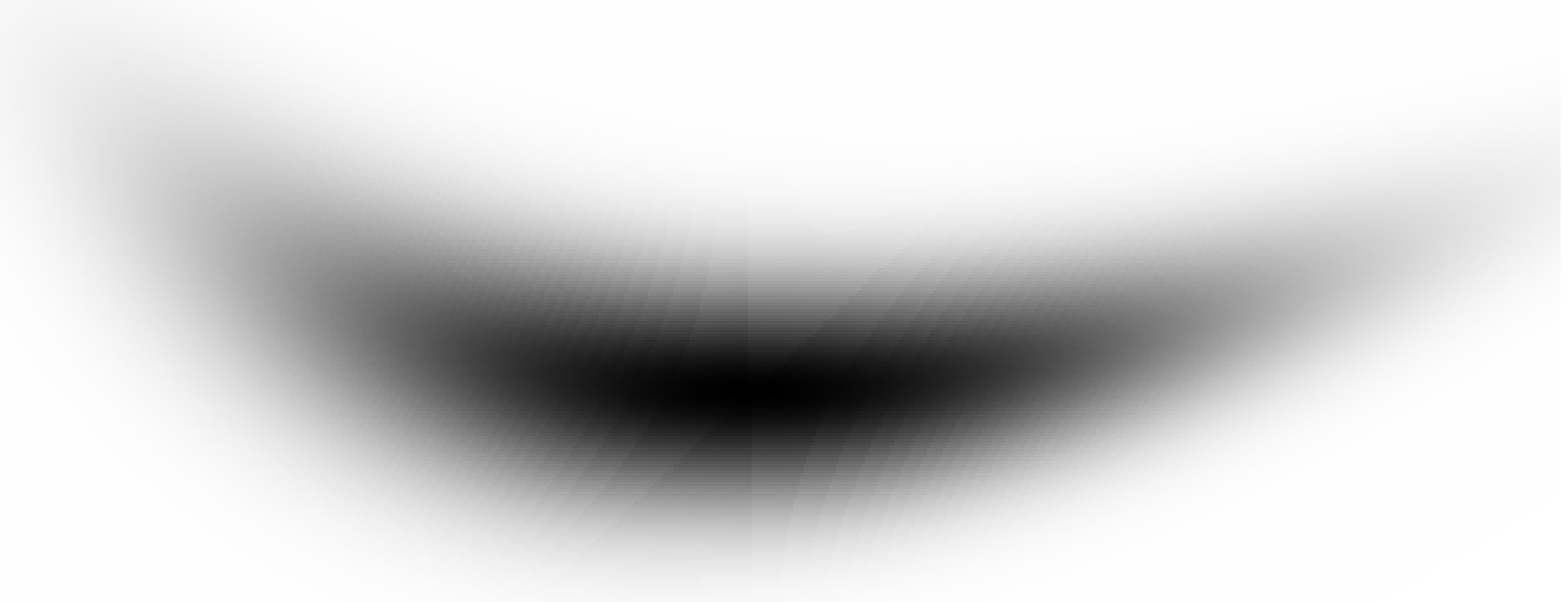,width=3.5cm}\psfig{figure=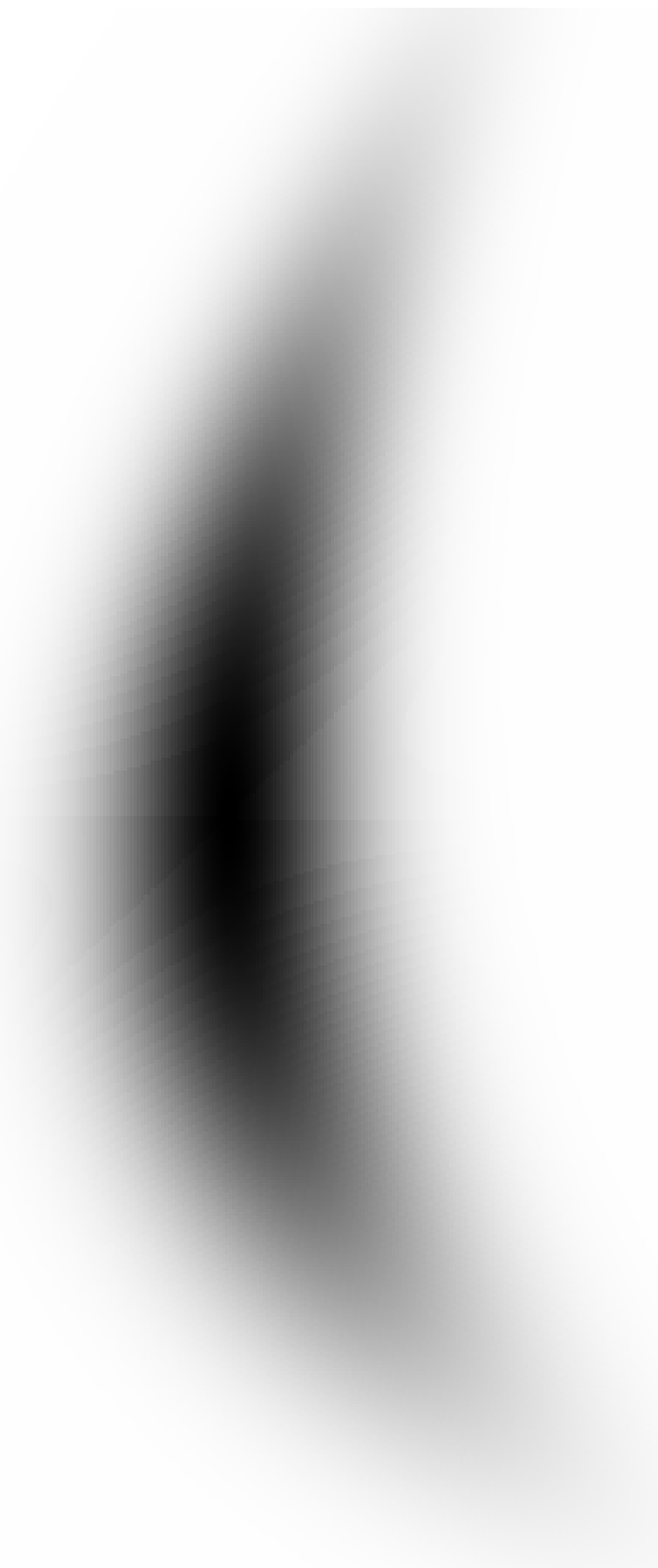,width=3.5cm}\\
\psfig{figure=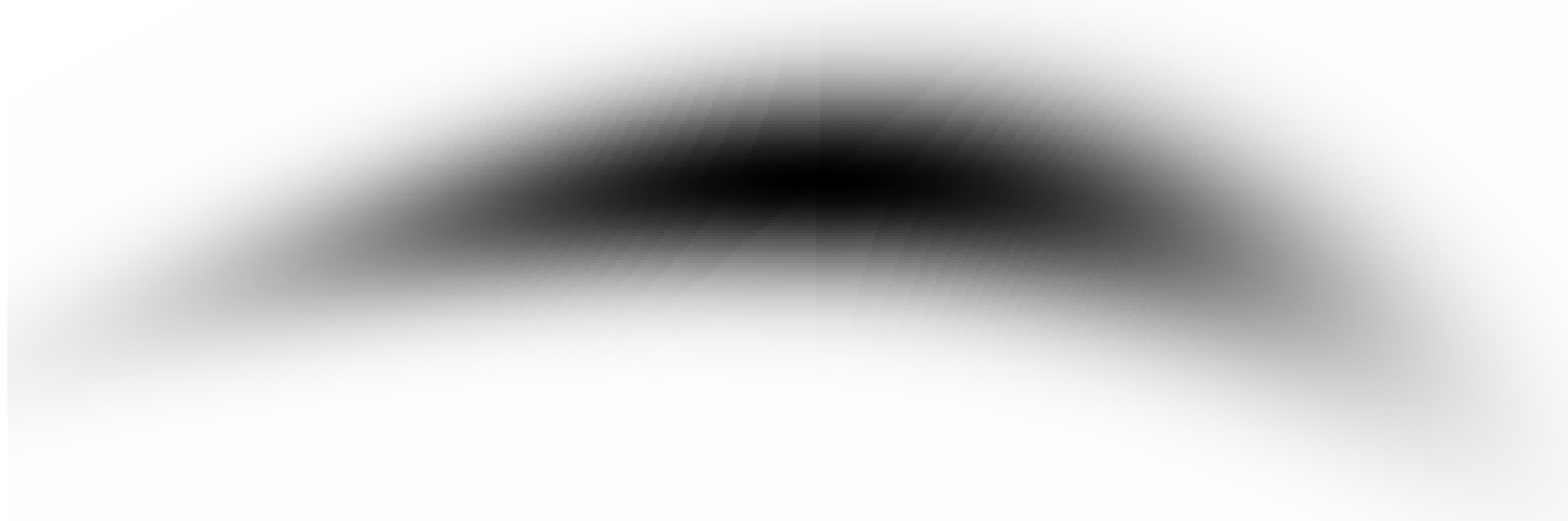,width=3.5cm}\psfig{figure=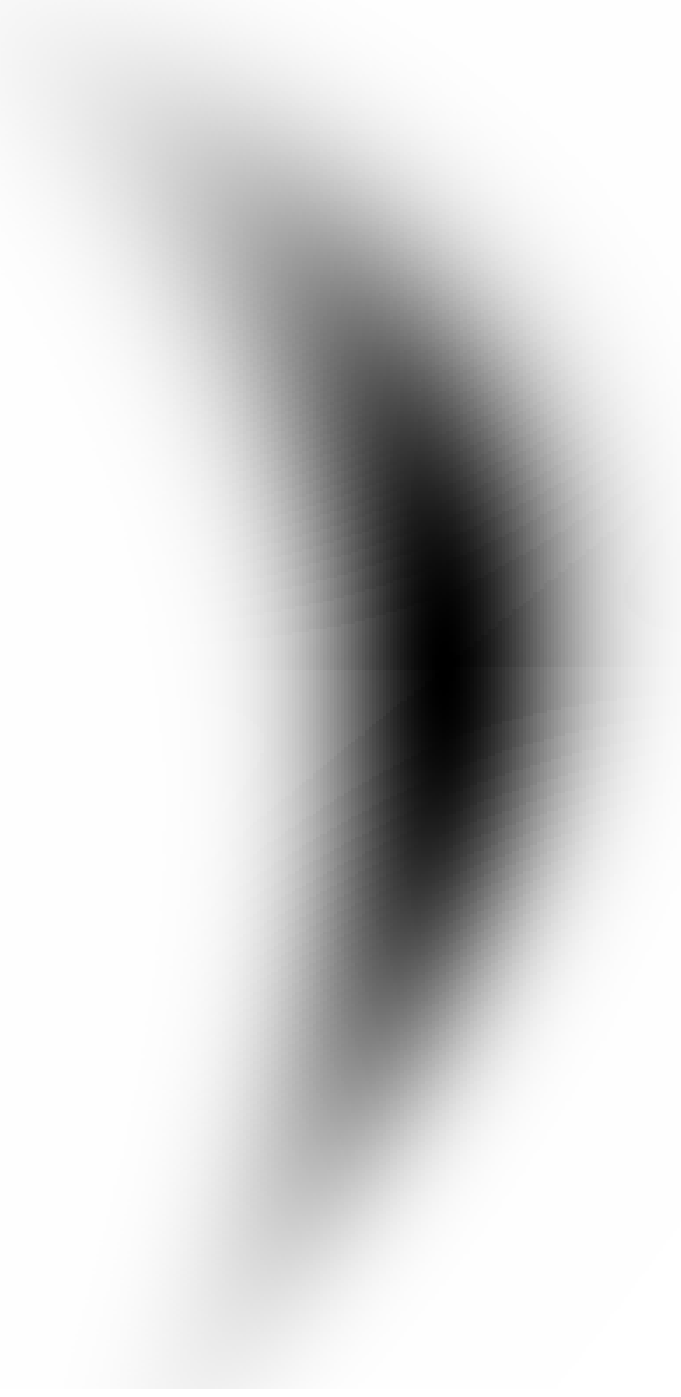,width=3.5cm}\\
\psfig{figure=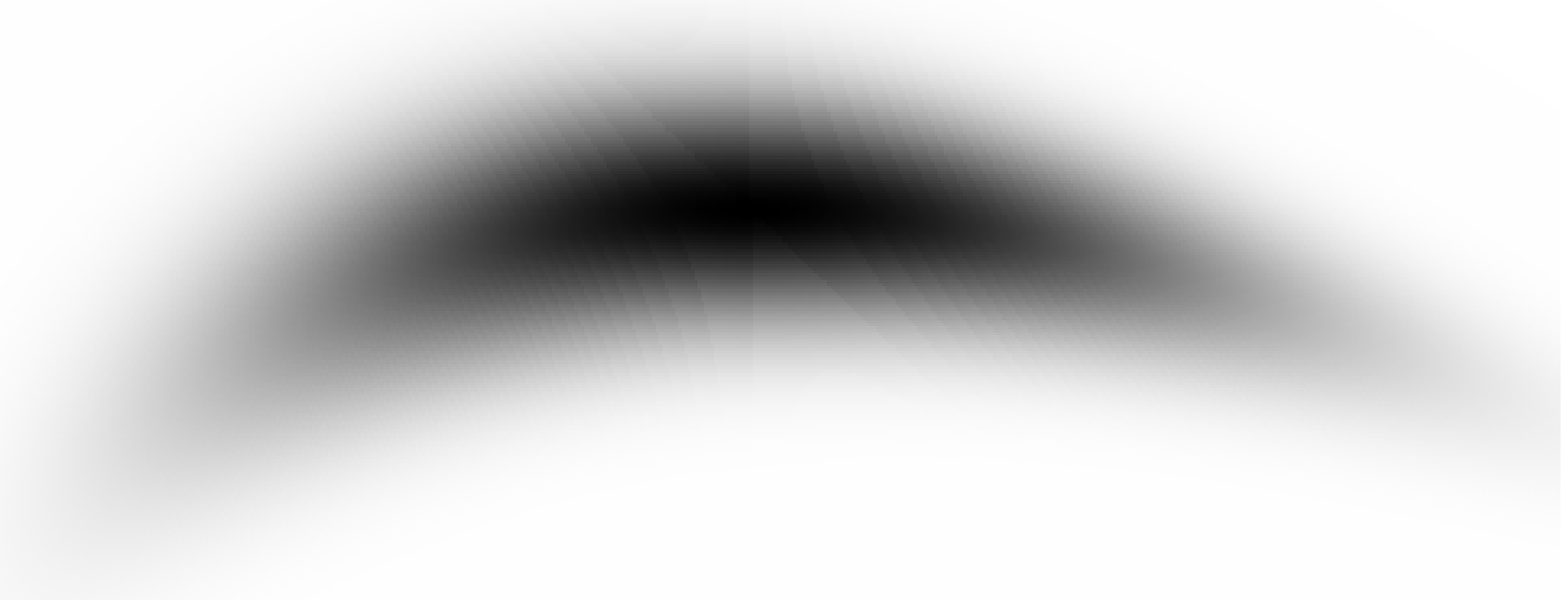,width=3.5cm}\psfig{figure=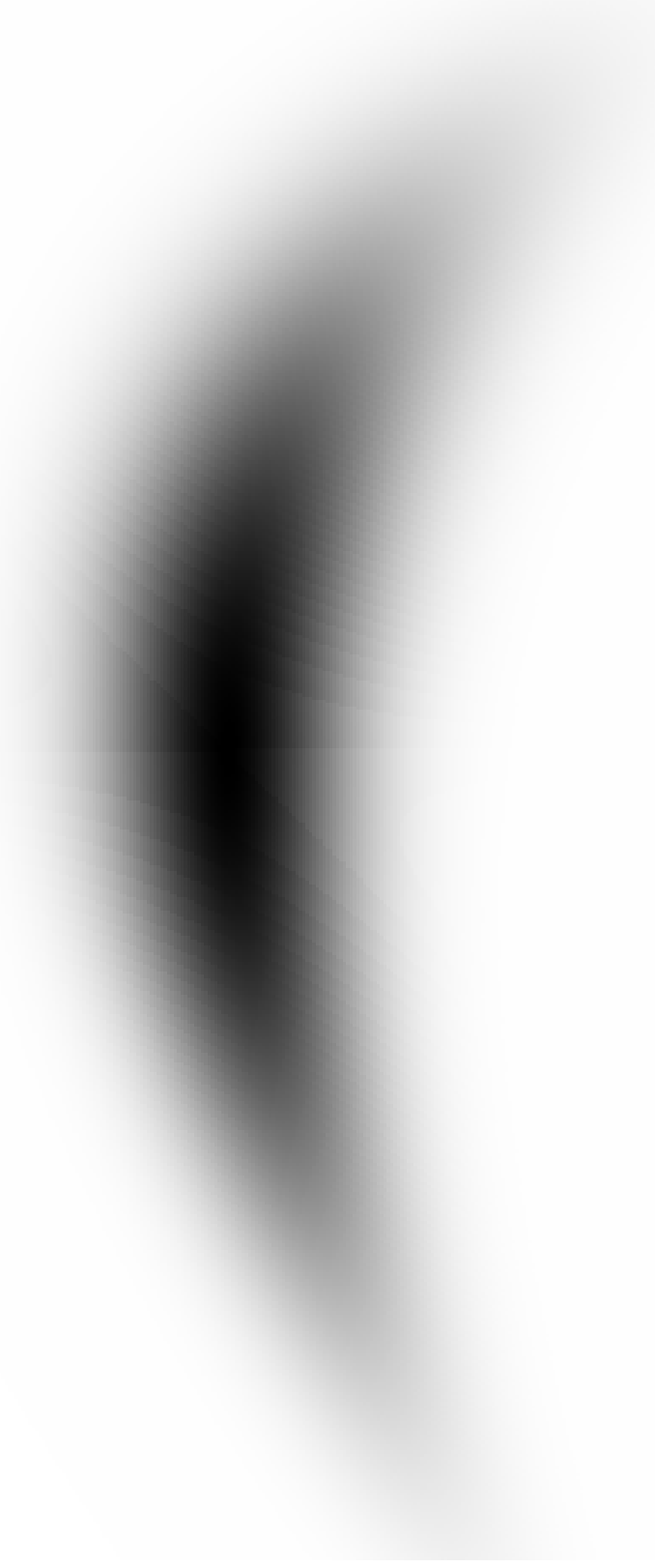,width=3.5cm}\\
\psfig{figure=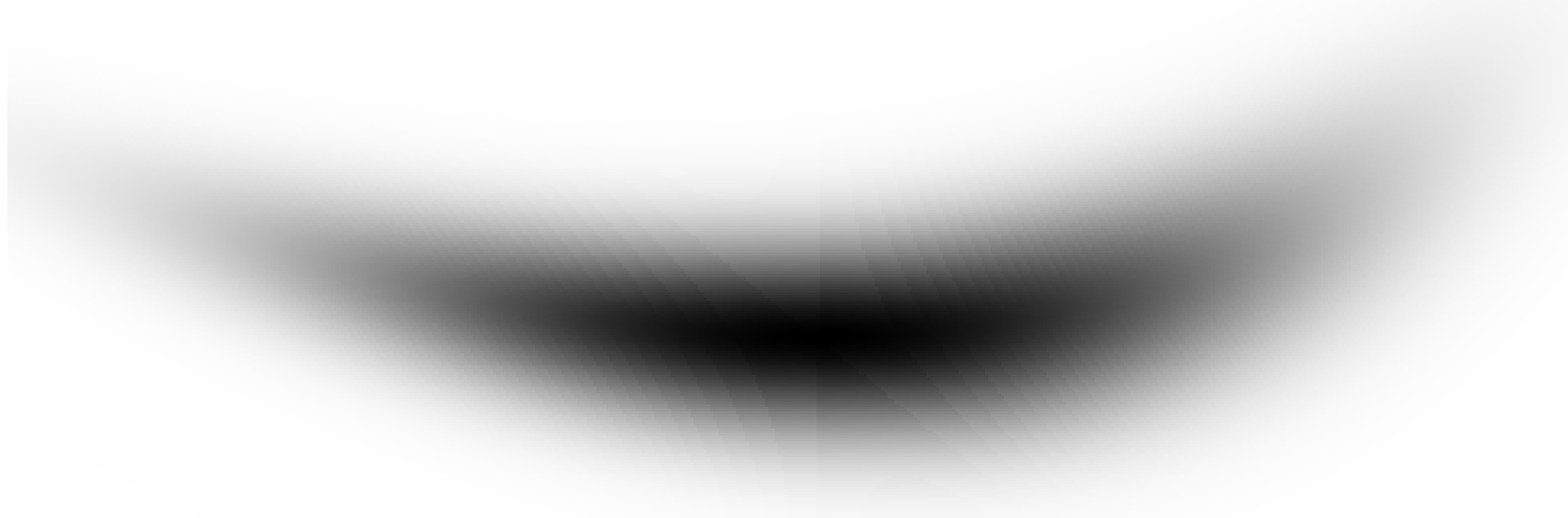,width=3.5cm}\psfig{figure=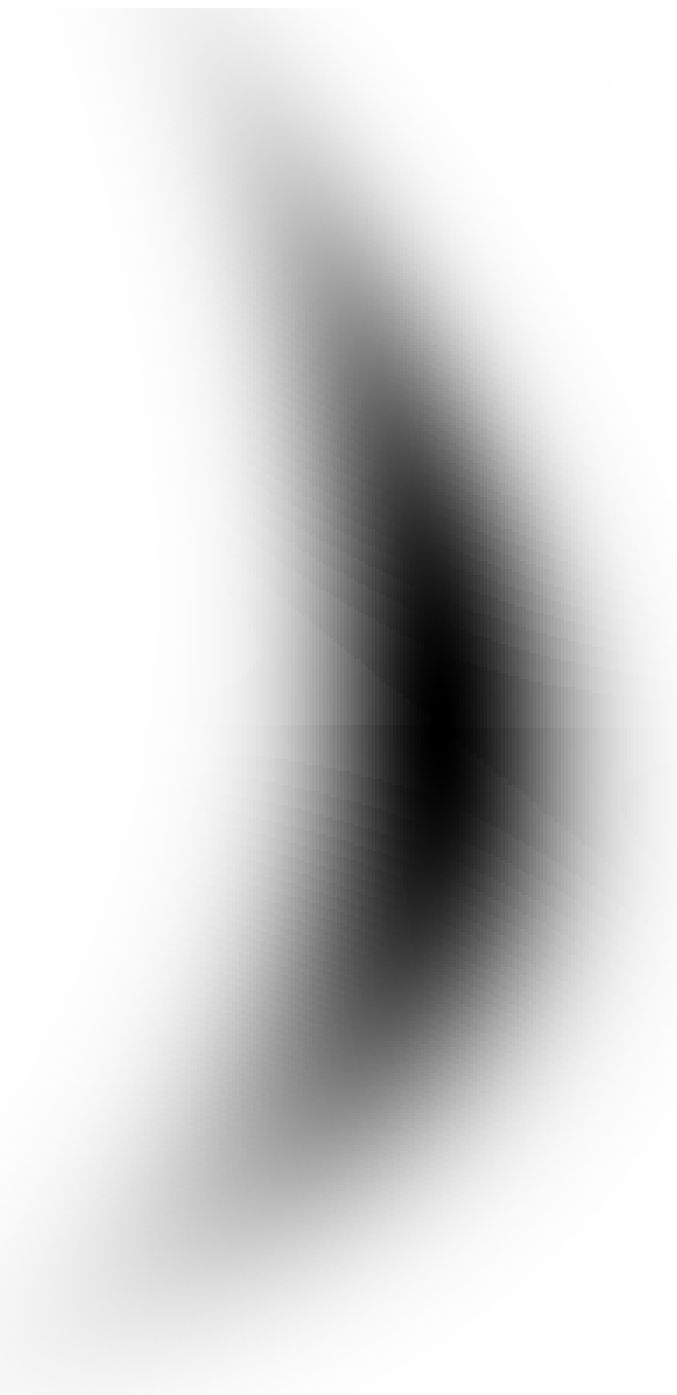,width=3.5cm}
\caption{Effect of twist on horizontally and vertically elliptical objects. Top panel: $T_1=1.3$ arcsec$^{-1}$ on a circular Gaussian with $\sigma=0.5"$. Second panel: $T_1=1.3$ arcsec$^{-1}$ on elliptical Gaussians $\sigma_{\rm major}=0.5"$, $e_1=0.9$ and $e_1=-0.9$. Third panel: $T_1=-1.3$ arcsec$^{-1}$ on the same elliptical Gaussians. Fourth panel: $T_2=1.3$ arcsec$^{-1}$; Bottom panel: $T_2=-1.3$ arcsec$^{-1}$. }
\label{fig:twist}
\end{figure}

Figure~\ref{fig:twist} shows the equivalent effect of twist. It might appear from this figure that twist has a different effect to turn, but a combination of $C_1$ and $C_2$ can achieve the same effect as $T_1$ or $T_2$; it is just that a pure $C_1$ distorts the object in a different direction to a pure $T_1$, for instance. Again, twist appears not to affect circular objects, and we will show this to be the case below. Its impact here is to turn horizontally or vertically elliptical objects into aerofoil shapes. Note the way in which $T$ components engage with the ellipticities to make aerofoils oriented in different directions. Positive $T_1$ or negative $T_2$ operate to twist horizontal objects into upward curving objects, with the front of the aerofoil pointing in opposite senses; negative $T_1$ or positive $T_2$ operate on horizontal objects to make downward curving objects. On the other hand, positive $T_1$ and positive $T_2$ bend vertical objects to the left, while negative $T_1$ and negative $T_2$ bend them to the right.

\subsection{Shapelet Space Behaviour}

We can gain insight into the behaviour of these distortion modes by examining their action in shapelet space. We use the polar shapelets of \citet{2002AJ....123..583B}, \citet{2003MNRAS.338...35R}, and \citet{2005MNRAS.363..197M}. As is described in the latter paper, polar shapelets can be described by their number of radial nodes $n$ and azimuthal nodes $m$, providing a basis set $|n,m\rangle$ for 2D localised objects. The shapelets require a length scale $\beta$ to be set, which is the standard deviation of the zeroth shapelet, a 2-D circular Gaussian. Then an image $|f\rangle$ is the sum of the shapelets with appropriate coefficients:
\begin{equation}
|f\rangle=\sum f_{nm} |n,m\rangle
\end{equation}
and a lensed image is the result of applying various operators to the source:
\begin{equation}
|f'\rangle=(1+\kappa \hat{K}+\rho\hat{R}+\gamma_i\hat{S}_i+F_i\hat{F}_i+G_i\hat{G}_i+T_i\hat{T}_i+C_i\hat{C}_i)|f\rangle
\end{equation}
where the terms are for convergence, rotation, shear, 1-flexion, 3-flexion, twist and turn respectively.
We wish to discover what these operators are in terms of the ladder operators which act on the basis:
\begin{eqnarray}
\hat{a}_r^\dagger |n,m\rangle &=& \sqrt{\frac{n+m+2}{2}}|n+1,m+1\rangle\nonumber\\
\hat{a}_r  |n,m\rangle&=& \sqrt{\frac{n+m}{2}}|n-1,m-1\rangle\nonumber\\
\hat{a}_l^\dagger  |n,m\rangle&=& \sqrt{\frac{n-m+2}{2}}|n+1,m-1\rangle\nonumber\\
\hat{a}_l  |n,m\rangle&=& \sqrt{\frac{n-m}{2}}|n-1,m+1\rangle .
\end{eqnarray}
The ladder operators obey commutation relations
\begin{eqnarray}
\left[\hat{a}_r,\hat{a}_r^\dagger\right]&=&1\nonumber\\
\left[\hat{a}_l,\hat{a}_l^\dagger\right]&=&1\nonumber
\end{eqnarray}
\begin{equation}
\left[\hat{a}_l,\hat{a}_r \right]=\left[\hat{a}_l,\hat{a}_r^\dagger\right]=\left[\hat{a}^\dagger_l,\hat{a}_r\right]=\left[\hat{a}_l^\dagger,\hat{a}_r^\dagger\right]=0.
\end{equation}
We can write position and derivative operators in terms of these ladder operators:
\begin{eqnarray}
\hat{x}&=&\frac{1}{2}\left[\hat{a}_r^\dagger+\hat{a}_l^\dagger+\hat{a}_l+\hat{a}_r\right]\nonumber\\
\hat{y}&=&\frac{i}{2}\left[\hat{a}_r^\dagger-\hat{a}_l^\dagger+\hat{a}_l-\hat{a}_r\right]\nonumber\\
\hat{\frac{\partial}{\partial x}}&=&\frac{1}{2}\left[-\hat{a}_r^\dagger-\hat{a}_l^\dagger+\hat{a}_l+\hat{a}_r\right]\nonumber\\
\hat{\frac{\partial}{\partial y}}&=&\frac{i}{2}\left[-\hat{a}_r^\dagger+\hat{a}_l^\dagger+\hat{a}_l-\hat{a}_r\right].
\label{eq:xop}
\end{eqnarray}
\begin{figure}
\hspace{-1.5cm}\psfig{figure=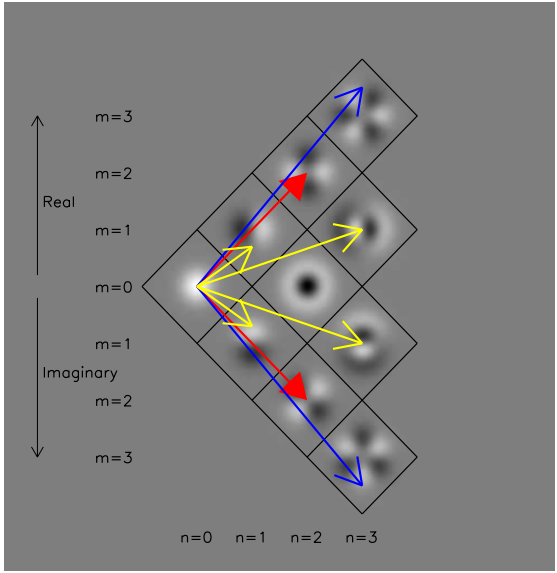,width=10cm}\vspace{-1cm}
\caption{Effect of shear and flexion on a circular Gausian in shapelet space. Shapelet profiles are displayed for the real part of polar shapelets in the top part of the figure, and for the imaginary part in the bottom part of the figure. Shear (red) takes power from the Gaussian shapelet $|0,0\rangle$ and places it in the spin-2 modes $|2,\pm2\rangle$. 1-flexion (yellow) moves power to spin-1 modes, while 3-flexion (blue) moves power to the spin-3 modes $|3,\pm3\rangle$.}
\label{fig:shearflexshap}
\end{figure}
Using these with equation (\ref{eq:map}), we can find forms for first and second order lensing operators. \citet{2005MNRAS.363..197M} and \citet{2007MNRAS.380..229M} have shown the forms for shear and flexion; these are summarised in figure~\ref{fig:shearflexshap}, which shows their effects on a circular Gaussian in shapelet space. Note that shear moves power from the $|0,0\rangle$ mode to the spin-2 modes $|2,\pm2\rangle$, while 1-flexion moves power to spin-1 modes, and 3-flexion moves power to the spin-3 modes $|3,\pm3\rangle$. 

We can carry out similar calculations for twist and turn, using equation (\ref{eq:map}) together with equation (\ref{eq:xop}). For twist, after routine but extensive non-commutative algebra we find 
\begin{eqnarray}
\hat{T}_1=-\frac{\beta}{8}\left[\left(1-\hat{a}_l^\dagger\hat{a}_l+\hat{a}_r^\dagger\hat{a}_r\right)\hat{a}_r(1-i)\right.\nonumber\\-\left(1-\hat{a}_r^\dagger\hat{a}_r+\hat{a}_l^\dagger\hat{a}_l\right)\hat{a}_l(1+i)\nonumber\\-\hat{a}_r^\dagger\left(\hat{a}_l^\dagger\hat{a}_l-\hat{a}_r^\dagger\hat{a}_r\right)(1+i)\nonumber\\\left.-\hat{a}_l^\dagger\left(\hat{a}_l^\dagger\hat{a}_l-\hat{a}_r^\dagger\hat{a}_r\right)(1-i)\right]\nonumber
\end{eqnarray}
\begin{eqnarray}
\hat{T}_2=-\frac{\beta}{8}\left[\left(1-\hat{a}_l^\dagger\hat{a}_l+\hat{a}_r^\dagger\hat{a}_r\right)\hat{a}_r(1+i)\right.\nonumber\\+\left(1-\hat{a}_r^\dagger\hat{a}_r+\hat{a}_l^\dagger\hat{a}_l\right)\hat{a}_l(1-i)\nonumber\\+\hat{a}_r^\dagger\left(\hat{a}_l^\dagger\hat{a}_l-\hat{a}_r^\dagger\hat{a}_r\right)(1-i)\nonumber\\\left.-\hat{a}_l^\dagger\left(\hat{a}_l^\dagger\hat{a}_l-\hat{a}_r^\dagger\hat{a}_r\right)(1+i)\right]
\end{eqnarray}
where the factor of $\beta$ takes into account the fact that the operators in equation (\ref{eq:xop}) work in units of $\beta$. One can consider what happens to a circular $(m=0)$ source operated on by e.g. $\hat{T}_1$; the third and fourth terms in the equation above for $\hat{T}_1$ vanish, as $\left(\hat{a}_l^\dagger\hat{a}_l-\hat{a}_r^\dagger\hat{a}_r\right)$ counts $m$; the first term initially acts with $a_r$ to move the state to a spin-1 state; then the term in brackets $\left(1-\hat{a}_l^\dagger\hat{a}_l+\hat{a}_r^\dagger\hat{a}_r\right)$ operates to give zero. The second term similarly gives zero, resulting in twist having no effect on circularly symmetric objects.

\begin{figure}
\hspace{-1.5cm}\psfig{figure=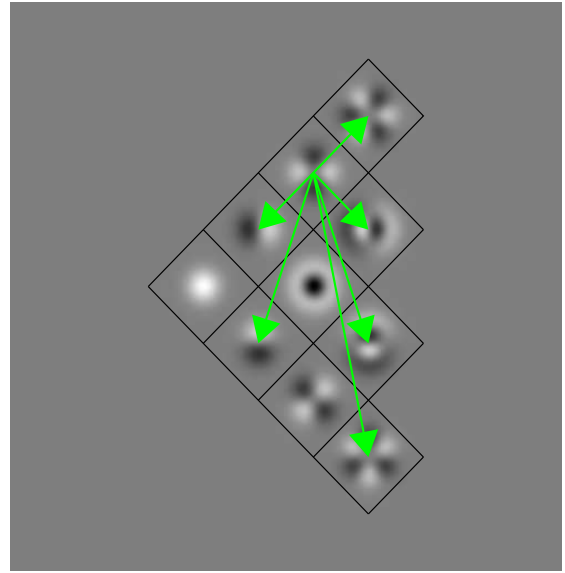,width=10cm}\vspace{-1cm}
\caption{Effect of twist in shapelet space. Circular Gaussians are not affected, so here we show the power moved from the $|2,2\rangle$ mode; note that twist pushes power into spin 1 and spin 3 modes.}
\label{fig:twistshap}
\end{figure}
The effect of these operators is shown in figure~\ref{fig:twistshap}. Since they have no impact for circular objects, we show the effect on the $|2,2\rangle$ mode. Note that power is moved to neighbouring spin 1 and spin 3 modes, with a rotation of $(1+i)$ or $(1-i)$ which gives the characteristic twisted form of the image.

\begin{figure}
\hspace{-1.5cm}\psfig{figure=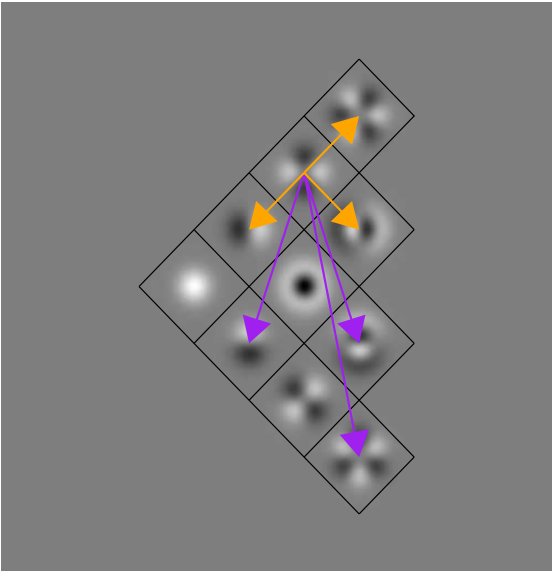,width=10cm}\vspace{-1cm}
\caption{Effect of turn in shapelet space. As for twist, circular Gaussians are not affected, so we again show the power moved from the $|2,2\rangle$ mode. The $C_1$ component (purple) pushes power into spin 1 and spin 3 modes on the opposite side of the diagram, while $C_2$ (orange) does the same on the near side of the diagram.}
\label{fig:turnshap}
\end{figure}

We carry out similar calculations for the equivalent turn description, again using operators given by equation (\ref{eq:xop}) together with the mapping equation (\ref{eq:map}). We find
\begin{eqnarray}
\hat{C}_1=\frac{i\beta}{8}\left[-\left(1+\hat{a}_l^\dagger\hat{a}_l-\hat{a}_r^\dagger\hat{a}_r\right)\hat{a}_l+\left(1+\hat{a}_r^\dagger\hat{a}_r-\hat{a}_l^\dagger\hat{a}_l\right)\hat{a}_r\right.\nonumber\\-\left.\hat{a}_l^\dagger\left(\hat{a}_l^\dagger\hat{a}_l-\hat{a}_r^\dagger\hat{a}_r\right)+\hat{a}_r^\dagger\left(\hat{a}_r^\dagger\hat{a}_r-\hat{a}_l^\dagger\hat{a}_l\right)\right]\nonumber
\end{eqnarray}
\begin{eqnarray}
\hat{C}_2=-\frac{\beta}{8}\left[-\left(1+\hat{a}_l^\dagger\hat{a}_l-\hat{a}_r^\dagger\hat{a}_r\right)\hat{a}_l-\left(1+\hat{a}_r^\dagger\hat{a}_r-\hat{a}_l^\dagger\hat{a}_l\right)\hat{a}_r\right.\nonumber\\+\left.\hat{a}_l^\dagger\left(\hat{a}_l^\dagger\hat{a}_l-\hat{a}_r^\dagger\hat{a}_r\right)+\hat{a}_r^\dagger\left(\hat{a}_r^\dagger\hat{a}_r-\hat{a}_l^\dagger\hat{a}_l\right)\right]
\end{eqnarray}
Here again we find that the impact on $m=0$ states is zero, using an identical argument to above. The effect of these operators is shown in figure~\ref{fig:turnshap}. Again we show the effect on the $|2,2\rangle$ mode; as for twist, power is moved to neighbouring spin 1 and spin 3 modes, but there are a different range of activated modes for $C_1$ and $C_2$, due to the different factors of $i$.

Now that we can describe twist/turn in shapelet space, we are in a position to construct practical estimators for measuring these quantities.

\section{Constraining Twist and Turn}\label{sect_estimators}

\subsection{Simple Estimators}

We can use the ladder operator form for twist to find a simple estimator for the new distortions. We consider the power that finishes in the $f_{11}$ component,
\begin{equation}
f_{11}'=f_{11}-\frac{\beta}{2}{\rm e}^{-i\pi/4} (T_1+i T_2) f_{22}
\label{eq:a11t1}
\end{equation}
where $f_{11}'$ is the component after twist. Since the mean untwisted $f_{11}$ is expected to be zero, we have the estimator
\begin{equation}
T_1^{\rm est}+iT_2^{\rm est}=-\frac{2}{\beta}{\rm e}^{i\pi/4}\frac{f_{11}}{f_{22}}
\label{eq:t1est}
\end{equation}
In a similar fashion, the turned $f_{11}$ coefficient is given by
\begin{equation}
f_{11}'=f_{11}+\frac{i\beta}{2\sqrt{2}}(C_1-i C_2) f_{22}
\label{eq:a11c1}
\end{equation}
so since the undistorted mean $f_{11}$ is expected to be zero, we obtain the estimator
\begin{equation}
C_1^{\rm est}-i C_2^{\rm est}= -\frac{2\sqrt{2}i}{\beta} \frac{f_{11}}{f_{22}}
\label{eq:c1est}
\end{equation}
Note the close relationship between the estimators for twist and turn, and the fact that they can indeed be written as superpositions of each other. 

However, it should be noted that these will only be pure estimators for twist/turn if 1-flexion is absent (or negligible). This is because 1-flexion also moves power into the $f_{11}$ mode (see \citet{2007MNRAS.380..229M}) in such a fashion that our estimator (e.g. for twist) is truly 

\begin{equation}
-\frac{2}{\beta}{\rm e}^{i\pi/4}\frac{f_{11}}{f_{22}}=T- \frac{F{\rm e}^{i\pi/4}}{2f_{22}}(3 f_{00}-3 f_{40})-\frac{F^* {\rm e}^{i\pi/4}}{4f_{22}}(\sqrt{2}f_{22} - 3 \sqrt{6}f_{42})
\end{equation}
The estimator is still of value to us despite this complication, as we wish to use it to see if there is a twist-like systematic in our survey; we now see that, at this shapelet order, systematic flexion can generate a twist-like effect. This simple estimator is therefore a test of combined second-order systematics, or of (real or systematic) twist on scales where flexion is negligible.

Nevertheless, it should be kept in mind that a pure estimation of twist will require a more extensive joint chi-squared fit of twist and flexion to several further orders of shapelets, in order to fully remove the degeneracy.

\subsection{Correction for Centroid Shift}

In addition, these simple estimators need correcting for the fact that twist/turn moves the centroid of the object. \citet{2005ApJ...619..741G} show that the centroid is moved by the $D$ tensor according to
\begin{eqnarray}
\Delta\bar{\theta}_1&=&-\langle\theta_1^2\rangle\left(\frac{3}{2}D_{111}+\frac{1}{2}D_{212}+\frac{1}{2}D_{221}\right)\nonumber\\&&-\langle\theta_1\theta_2\rangle\left(D_{112}+D_{121}+D_{222}\right)-\langle\theta_2^2\rangle\frac{1}{2}D_{122}\nonumber\\
\Delta\bar{\theta}_2&=&-\langle\theta_1^2\rangle\frac{1}{2}D_{211}-\langle\theta_1\theta_2\rangle\left(D_{221}+D_{212}+D_{111}\right)\nonumber\\&&-\langle\theta_2^2\rangle\left(\frac{3}{2}D_{222}+\frac{1}{2}D_{121}+\frac{1}{2}D_{112}\right)
\end{eqnarray}
where we have written a form which assumes less symmetry than Goldberg \& Bacon; this is necessary for our generalised $D$ tensor. Putting the values of the $D$ tensor, equation (\ref{eq:d}), into this equation we find
\begin{equation}
\Delta\bar{\theta}_1+i\Delta\bar{\theta}_2=\frac{R^2}{4\beta}\left[6F+5F^*e+Ge^*+iC^*e+(i-1)Te\right]
\end{equation}
where $R^2$ is the size quadrupole and $e$ is the unweighted ellipticity as given in \citet{2007MNRAS.380..229M}. Note again that for circular $(e=0)$ objects, twist and turn have no effect.

\citet{2007MNRAS.380..229M}  showed that the effect of the centroid shift is to alter observable flexion estimators $\hat{F}_1+i\hat{F}_2$ by substracting a term $(\Delta \bar{\theta} \hat{D}_r+\Delta\bar{\theta}^*\hat{D}_l)$, and the same applies here. The shift operators $\hat{D}$ are given by
\begin{equation}
\hat{D}_r=\frac{1}{2}\left(a_r^\dagger-a_l \right) \hspace{1cm} \hat{D}_l=\frac{1}{2}\left(a_l^\dagger-a_r \right)
\end{equation}
Hence we find that equation (\ref{eq:a11t1}) becomes
\begin{eqnarray}
f_{11}'&=&f_{11}-\frac{\beta}{2}{\rm e}^{-i\pi/4}T f_{22}\nonumber\\&-&\frac{R^2}{8\beta}T\left[(i-1)(e_1+ie_2)(f_{00}-f_{20})\right.\nonumber\\&&+\left.\sqrt{2}(i+1)(e_1-ie_2)f_{22}\right]
\end{eqnarray}
By dividing both sides by $f_{22}$, we can therefore propose the corrected estimator
\begin{equation}
T^{\rm est}=-\frac{2}{\beta}{\rm e}^{i\pi/4}\frac{f_{11}}{f_{22}}\left(1-\left\langle\frac{R^2(e_1+ie_2)(f_{00}-f_{20})}{2\sqrt{2}\beta^2f_{22}} \right\rangle\right)^{-1}
\label{eq:t1estcor}
\end{equation}
If we label the term in brackets as $B$, we similarly find for the turns
\begin{equation}
C^{\rm est}=-\frac{2\sqrt{2}i}{B\beta}\frac{f_{11}}{f_{22}}
\label{eq:c2estcor}
\end{equation}
So our corrected estimators differ from our naive estimators only by a factor of $B$. Again, twist can be written as a superposition of turns, and vice versa. As in the previous section, the presence of flexion would mean that these are not pure estimators of twist; the estimators should again be interpreted as showing combined second-order systematics, or (real or systematic) twist on scales where flexion is negligible.

\section{First Measurements with Twist/Turn Estimators}\label{sect_measurement}

We are now in a position to measure these twist/turn estimators on real data. We use the STAGES mosaic observed with the Hubble Space Telescope \citep{2007AAS...21113220G, 2008MNRAS.385.1431H}. This is a 0.25 square degree field observed with the Advanced Camera for Surveys in the F606W band, covering 80 ACS tiles in 80 orbits. Drizzling is used to obtain an effective pixel size of $0.03"$.

We use the same galaxy catalogue as \citet{2008MNRAS.385.1431H}, deconvolving and decomposing all objects into shapelets using the methods developed in \citet{2003MNRAS.338...35R, 2003MNRAS.338...48R, 2005MNRAS.363..197M}. The analysis will be described in full in Bacon et al (2009); we obtain a shapelet catalogue for 56,000 galaxies, together with measures of $\beta$ and $R^2$ for all objects. The shapelets are normalised so that $f_{00}=1$.

\begin{figure}
\psfig{figure=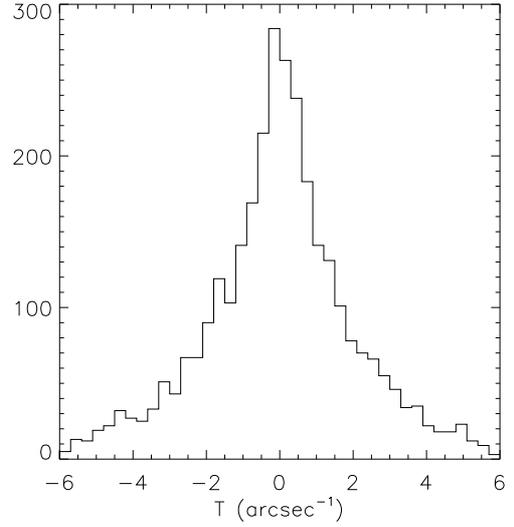,width=8cm}
\caption{Histogram of twist estimator values in one component in the STAGES survey, for objects with F606 magnitude$<23.5$, $|T| < 6.0$ arcsec$^{-1}$ and $\beta > 1$ pixel.}
\label{fig:twisthist}
\end{figure}

We measure the twist/turn estimators for objects with F606 magnitude $<23.5$ using equations (\ref{eq:t1estcor}) to (\ref{eq:c2estcor}); we find that for this sample, $B=-1.59\pm0.01$. Since the twist and turn measurements are equivalent, here we choose to present results in terms of twist. The histogram of twist estimators is shown in figure~\ref{fig:twisthist}; this includes $3\sigma$ cuts for outliers with $|T|>6.0$~arcsec$^{-1}$, and we only consider objects with $\beta>$1 pixel to avoid oversampling. The first thing to note is that our twist estimator is more noisy than shear and flexion estimators, having a standard deviation in one component of 2.0 arcsec$^{-1}$; much of this noise is due to intrinsic shape variance of the objects. The turn estimator as defined has a larger standard deviation of 2.9 arcsec$^{-1}$.

We find mean values over the STAGES survey of $\bar{T}_1=-0.016\pm0.036$ arcsec$^{-1}$, $\bar{T}_2=-0.009\pm0.037$ arcsec$^{-1}$. These are consistent with zero, as we might hope for a systematic mode. At present the constraint is fairly weak, as gravitational flexion signals are at the level of 0.001 to 0.01 arcsec$^{-1}$; however, in upcoming lensing surveys the much larger area will lead to twist/turn constraints at the $10^{-4}$ level, which will provide important checks on systematics.

\begin{figure}
\psfig{figure=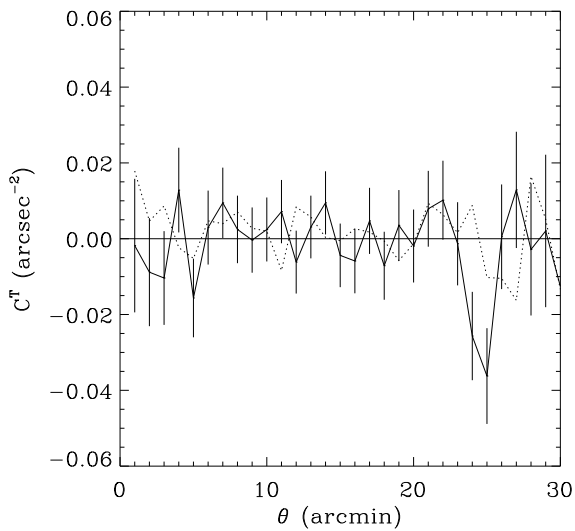,width=8cm}
\caption{STAGES twist estimator correlation function. Solid line: $\langle T_1 T_1 \rangle$ correlation function. Dashed line: $\langle T_2 T_2 \rangle$ correlation function. Dotted line: $\langle T_1 T_2 \rangle$ correlation function.}
\label{fig:twistcor}
\end{figure}

We can further explore whether the twist/turn estimator is activated as a systematic in the STAGES survey by measuring its correlation functions. As with shear correlation functions, the twists should be rotated before they are correlated; however, while shear has to be rotated by factors of e$^{i2\phi}$ where $\phi$ is the position angle of the line joining a pair of objects, twist has to be rotated by factors of e$^{i\phi}$ on account of its vector nature:
\begin{eqnarray}
T_1^{\rm rot}=T_1 \cos{\phi}+T_2 \sin{\phi} \nonumber\\T_2^{\rm rot}=-T_1 \sin{\phi}+T_2 \cos{\phi} 
\end{eqnarray}
We can then construct correlation functions
\begin{eqnarray}
C^{T}_{11}(\theta)=\langle T_1^{\rm rot}(\vec{\theta}_i) T_1^{\rm rot}(\vec{\theta}_i+\vec{\theta})\rangle\nonumber\\
C^{T}_{22}(\theta)=\langle T_2^{\rm rot}(\vec{\theta}_i) T_2^{\rm rot}(\vec{\theta}_i+\vec{\theta})\rangle\nonumber\\
C^{T}_{12}(\theta)=\langle T_1^{\rm rot}(\vec{\theta}_i) T_2^{\rm rot}(\vec{\theta}_i+\vec{\theta})\rangle
\end{eqnarray}
We have measured these correlation functions for twist estimators in STAGES, and display the results in figure~\ref{fig:twistcor}. Here error bars are estimated by $\sigma^2/\sqrt{N_{\rm pairs}}$ where $\sigma$ is the standard deviation of twist and $N_{\rm pairs}$ is the number of galaxy pairs in a bin. We find that the correlation functions are almost all consistent with zero signal, with reduced $\chi^2_\nu={0.87, 0.53, 0.39}$ for $C^T_{11}, C^T_{22}, C^T_{12}$ respectively. 

\section{Conclusion}\label{sect_summary}

In this paper, we have written down for the first time the full weak image distortion relevant to weak lensing, to second order. This involved the discovery of a new image distortion, the twist, which can be written in an alternative form as the turn.

We reviewed weak lensing distortions to first order, recouching the lens mapping in terms of Pauli matrices. We noted the existence of a non-gravitational mode, the rotation. This sets a precedent which we also see at second order. 

We then extended the formalism to second order; at this point it became clear that the gradient of the rotation gives a new mode, which we call the turn. A further orthogonal mode in $D$ was found by seeking a combination of Pauli matrices orthogonal to all known modes; this new mode was called the twist. We showed that twist and turn can be written in terms of each other regarding their observational consequences, and were then able to write down the full image distortion mapping to second order.

We explored the properties of twist/turn, finding that it is a vector quantity. Its visual effect was shown, as was its impact in shapelet space. With the ladder operator formalism we showed that twist/turn has no effect on circularly symmetric objects, but only objects with non-zero ellipticity. We saw how twist/turn moves power from spin-2 modes to modes with spin-1 and spin-3.

Using our ladder operator forms for twist and turn, we found simple estimators for the distortions; however, twist/turn causes a centroid shift which needs to be taken into account. This leads to the inclusion of a common factor $B$ in the estimators for twist and turn. In addition, any flexion present would contribute to our simple estimator, which should therefore be used either as a means to detect any second-order systematic, or to measure real or systematic twist on scales where flexion is negligible.

We used these estimators to constrain twist for the first time in the HST STAGES survey. We noted that our estimator has a larger intrinsic noise scatter compared to 1- or 3-flexion, but its mean value across the survey is already at an interesting level for checking large flexion systematics. We found that mean twist is consistent with zero in this survey. We also measured twist correlation functions, and found that they too were consistent with zero.

The two quantities introduced in this study complete the set of distortions to second order. They will be of use in testing for systematic effects, and have a certain elegance of their own.

\section{Acknowledgements}
We would like to thank Kathy Bacon, Xinzhong Er, Meghan Gray, Alan Heavens, Catherine Heymans, Barnaby Rowe, Peter Schneider, Uro\v{s} Seljak and Andy Taylor for useful discussions, and the STAGES collaboration for use of their data. We thank the referee for very useful comments. DB is funded by an STFC Advanced Fellowship and an RCUK Research Fellowship.

\bibliographystyle{mn2e}
\bibliography{aamnem,references}

\end{document}